\documentclass[journal]{IEEEtran}

\UseRawInputEncoding
\usepackage{blindtext}
\usepackage{soul} 
\usepackage{cite}
\usepackage{amsmath,amssymb,amsfonts}
\usepackage{float} 
\usepackage[noend]{algpseudocode}
\usepackage{graphicx}
\usepackage{textcomp}
\usepackage[linesnumbered,ruled,vlined]{algorithm2e}
\usepackage{svg}
\usepackage{xcolor}
\usepackage{lipsum}
\definecolor{my_brown}{RGB}{115,115,115}
\begin{document}

\title{Deep Reinforcement Learning for EH-Enabled Cognitive-IoT Under Jamming Attacks}

\author{Nadia~Abdolkhani,~\IEEEmembership{Student Member,~IEEE,}
        Nada~Abdel~Khalek,~\IEEEmembership{Student Member,~IEEE,}
        and~Walaa~Hamouda,~\IEEEmembership{Senior~Member,~IEEE}

\thanks{N. Abdolkhani, N. Abdel Khalek, and W. Hamouda are with the Department of Electrical and Computer Engineering, Concordia University, Montreal, QC, H3G 1M8, Canada (e-mail: n\_abdolk@ece.concordia.ca; n\_abdelk@ece.concordia.ca; hamouda@ece.concordia.ca).}
}


\maketitle

\begin{abstract}
In the evolving landscape of the Internet of Things (IoT), integrating cognitive radio (CR) has become a practical solution to address the challenge of spectrum scarcity, leading to the development of cognitive IoT (CIoT). However, the vulnerability of radio communications makes radio jamming attacks a key concern in CIoT networks. In this paper, we introduce a novel deep reinforcement learning (DRL) approach designed to optimize throughput and extend network lifetime of an energy-constrained CIoT system under jamming attacks. This DRL framework equips a CIoT device with the autonomy to manage energy harvesting (EH) and data transmission, while also regulating its transmit power to respect spectrum-sharing constraints. We formulate the optimization problem under various constraints, and we model the CIoT device's interactions within the channel as a model-free Markov decision process (MDP). The MDP serves as a foundation to develop a double deep Q-network (DDQN), designed to help the CIoT agent learn the optimal communication policy to navigate challenges such as dynamic channel occupancy, jamming attacks, and channel fading while achieving its goal. Additionally, we introduce a variant of the upper confidence bound (UCB) algorithm, named UCB-IA, which enhances the CIoT network's ability to efficiently navigate jamming attacks within the channel. The proposed DRL algorithm does not rely on prior knowledge and uses locally observable information such as channel occupancy, jamming activity, channel gain, and energy arrival to make decisions. Extensive simulations prove that our proposed DRL algorithm that utilizes the UCB-IA strategy surpasses existing benchmarks, allowing for a more adaptive, energy-efficient, and secure spectrum sharing in CIoT networks.
\end{abstract}


\IEEEpeerreviewmaketitle

\section{Introduction}

\IEEEPARstart{I}{n} the midst of the rapid proliferation of Internet of Things (IoT) devices and services, IoT networks grapple with spectrum scarcity and congestion challenges. Given that IoT lacks dedicated bandwidth and shares the spectrum with various RF technologies like Bluetooth, addressing these challenges becomes imperative \cite{nada_survey_2023}. Cognitive radio (CR) technology has emerged as a strategic solution that allows IoT devices to efficiently share the spectrum with licensed users, or primary users (PUs), thereby improving the spectral utilization efficiency \cite{Tanab_Survey_2017}. There are three main CR access approaches by which secondary users (SUs) or unlicensed users can access the spectrum, which are the interweave, underlay, and overlay approaches. Using the interweave approach, SUs can only access the licensed spectrum if and only the PUs are not utilizing it \cite{nada_icc22}. On the other hand, in the underlay approach both the SUs and PUs coexist provided that the SUs do not interfere with the PUs \cite{IoT_2024_Nada_Nadia}. Finally, in the overlay approach, the SUs assist the PUs in exchange for spectrum access privileges \cite{Han_Xu_Jin_Li_Chen_Fang_Xu_2022}.

To effectively share the spectrum in underlay CR, cognitive IoT (CIoT) devices have to dynamically adjust their transmission power to abide by the interference constraint set by the primary network while also aiming to maximize their throughput.  Effective power control is essential, especially for energy-constrained CIoT devices, which must not only focus on maximizing their throughput but also on extending the network's lifetime. Energy harvesting (EH) offers a solution for CIoT devices to overcome their energy constraints, allowing them to be \textit{self-sufficient} and ensuring network continuity without the necessity for external sources. EH involves harnessing energy from renewable sources in the environment, such as solar, wind, and radio frequency (RF) signals. Among these sources, RF signals offer a particularly promising future due to their predictable, stable nature, and cost-effectiveness \cite{Guo_Yang_Zou_Qian_Zhu_Hanzo_2019, Liu_Xia_Hu_Zheng_Zhang_2022}. In RF EH, CIoT devices can gather energy from RF signals produced by nearby devices to recharge their batteries \cite{Umeonwuka_Adejumobi_Shongwe_Thokozani_2024}. That is, EH RF allows CIoT devices to operate completely untethered, eliminating the need for support from a wired power source or frequent battery replacements.

Although CIoT networks benefit from the opportunistic spectrum access, they are vulnerable to jamming attacks given the inherent shared and broadcast nature of radio wave propagation \cite{Lin_Qiu_Wang_Zhang_2024}. As indicated by \cite{pirayesh2022jamming}, the ease of launching jamming attacks, thanks to the advancements in software-defined radio, underscores the urgency of securing wireless networks against both intentional and unintentional jamming. In the presence of such attacks, the jammer can disrupt communication channels by introducing continuous jamming signals or brief jamming pulses across single or multiple frequency bands, leading to a degradation in the signal-to-noise ratio (SNR) \cite{pirayesh2022jamming}. Consequently, the throughput capacity of ongoing transmissions diminishes, and in some cases, the transmission may be completely interrupted \cite{Xu_Lou_Zhang_Sang_2020}. Therefore, jamming attacks pose a significant threat to CIoT transmissions, especially under energy and spectrum sharing constraints. Hence, there is a need for robust and intelligent countermeasures that not only enhance the performance of CIoT networks but also ensure their security and prolong their lifetime.

\section{Background and Motivation}
\subsection{Related Works}
Game theory has been employed in the literature for anti-jamming communication strategies \cite{8558506, 9380308,9172090,9200488}. Jia \textit{et al.} \cite{8558506} explored anti-jamming channel selection strategies in dense wireless networks under the presence of jammers, framing the anti-jamming communication problem as a Stackelberg game. A multi-domain anti-jamming strategy, considering both time and power domains, was proposed in \cite{9200488}. Using a Stackelberg game model, a genetic algorithm was developed to determine the optimal frequency-hopping speed and transmit powers. The Colonel Blotto anti-jamming game in \cite{9380308} offers a defense strategy for slow fading IoT networks using orthogonal frequency-division multiplexing channels. This approach minimizes the worst-case jamming effect and optimizes power allocation, effectively combating jamming attacks while conserving IoT sensors' power.

In multi-user networks, Xu \textit{et al.} investigated a game-theoretic anti-jamming spectrum access approach \cite{9172090}. The authors develop an internal coordination-external confrontation game model and design a coordination-learning-based anti-jamming spectrum access algorithm to achieve correlated equilibrium. While game-theoretical strategies have been explored for anti-jamming communications, they often rely on impractical prerequisites such as prior knowledge of the jamming pattern and may prove ineffective when confronted with constantly evolving jamming tactics \cite{8094301}. Consequently, the utilization of intelligent algorithms in anti-jamming communications is encouraged due to their increased unpredictability for jammers, as these algorithms dynamically adjust to the spectrum's state \cite{Wang_Wang_Xu_Chen_Jia_Liu_Yang_2020}.

By employing reinforcement learning (RL) algorithms, jamming behaviors can be learned through a process of ``trial-and-error" with the environment, even without access to specific information about the jammer. Works such as \cite{Ali_Naser_Muhaidat_2023,Shen_Wang_Jin_Zhang_2021,Aref_Jayaweera_2019_int_av,Han_Xu_Jin_Li_Chen_Fang_Xu_2022,Zhou_Li_Niu_Qin_Zhao_Wang_2020,Bi_Wu_Hua_2019} have addressed jamming attacks in the spectrum domain by using learning-based frequency hopping strategies. Ali \textit{et al.} proposed an energy-efficient Dueling deep Q-network (DQN) for implementing an anti-jamming frequency hopping strategy in CIoT networks \cite{Ali_Naser_Muhaidat_2023}. Shen \textit{et al.} employed a convolutional double deep Q-network (DDQN) for smart channel selection to counter jamming attacks \cite{Shen_Wang_Jin_Zhang_2021}. Aref \textit{et al.} proposed a convolutional DDQN framework to address interference and jamming in wideband spectrums, focusing on reducing computational complexity \cite{Aref_Jayaweera_2019_int_av}.

To simulate the communication environment of an overlay CR network under indiscriminate jamming attacks, Han \textit{et al.} utilized a generative adversarial network (GAN) \cite{Han_Xu_Jin_Li_Chen_Fang_Xu_2022}. Subsequently, a soft actor-critic deep reinforcement learning (DRL) approach, incorporating convolutional layers, was used to determine the optimal spectrum access policy within the simulated environment. Zhou \textit{et al.} tackled the issue of anti-jamming by using a collaborative multi-agent reinforcement learning (MARL) approach based on Q-learning for channel selection under intelligent dynamic jamming \cite{Zhou_Li_Niu_Qin_Zhao_Wang_2020}. Bi \textit{et al.} introduced a frequency hopping strategy for multi-user CR networks that utilizes a DDQN framework \cite{Bi_Wu_Hua_2019}. The base station takes charge of training the DDQN to identify the most effective channel selection strategy, which is subsequently deployed to the CR users post-training.

Works such as \cite{Aref_Jayaweera_2019,Li_Xu_Xu_Liu_Wang_Li_Anpalagan_2020,Liu_Li_Cui_Liu_Chen_Chen_Li_Xu_2023} have focused attention towards refining anti-jamming communication strategies that encompass not only frequency selection but also sub-band selection within each frequency. In \cite{Aref_Jayaweera_2019}, a convolutional DDQN was designed for anti-jamming communications within a heterogeneous wideband spectrum. This framework determines both the selection of sub-bands for access and the choice of channels within those sub-bands. Employing multitask transfer learning, the authors accelerate training by treating each sub-band as a distinct task. Li \textit{et al.} introduced a hierarchical DRL algorithm \cite{Li_Xu_Xu_Liu_Wang_Li_Anpalagan_2020}. This algorithm utilizes two convolutional DQNs, one for determining the frequency band and another for selecting the specific frequency within the chosen band, prioritizing frequencies with the lowest likelihood of being jammed. Liu \textit{et al.} introduced an anti-jamming approach for both the time and frequency domains in cellular CR networks \cite{Liu_Li_Cui_Liu_Chen_Chen_Li_Xu_2023}. The approach utilizes a convolutional DQN with a long short-term memory (LSTM) layer and enables the CR transmitter to decide on transmission timing in the time domain and conduct joint channel-bandwidth selection in the frequency domain.

The above approaches to countering jamming attacks relied on the assumption that multiple channels are available and jammers can simultaneously target all channels. As a result, SUs may select channels less likely to be jammed or switch to an alternative channel upon encountering jamming. However, there are scenarios where SUs are configured to opportunistically transmit data on a specific channel, such as when permitted by the licensed channel holder. Therefore, power control approaches have been explored in wireless communication to counter jamming attacks. Notably, power control schemes in \cite{9778818,Chen_Li_Xu_Xiao_2018} involve assessing channel conditions and adapting transmit power to counteract jamming. Geng \textit{et al.} developed a dynamic anti-jamming model using Q-learning to determine the optimal anti-jamming power, where users lack knowledge of the game model \cite{9778818}. Despite being employed for power control, Q-learning exhibits extended convergence times to achieve the optimal strategy when faced with a considerable number of states, and it may occasionally fail to converge \cite{Chen_Li_Xu_Xiao_2018}.

Deep reinforcement learning (DRL), which combines both RL and deep learning (DL), has been utilized to overcome the challenges of Q-learning. Nguyen \textit{et al.} introduced a method based on a DDQN to acquire an effective communication policy, encompassing channel access and transmission power adjustments to address various jamming scenarios \cite{Nguyen_Nguyen_Do_2021}. Similarly, Xu \textit{et al.} use a transformer encoder-based DDQN for channel and transmit power selection by a secondary transmitter under the presence of a jammer \cite{Xu_Lou_Zhang_Sang_2020}. Ali \textit{et al.} used clear channel assessment data to train a DDQN to learn an effective anti-jamming strategy for a CR agent that dynamically switches channels and selects the appropriate transmit power \cite{Ali_Lunardi_2022}. Aref \textit{et al.} proposed a multitask DQN architecture for multi-agent environments to maintain the required quality of service (QoS) by adjusting the transmit power and frequency hopping in a wideband spectrum \cite{Aref_Jayaweera_2021}. Chen \textit{et al.} proposed a convolutional DQN for power control in CIoT networks under the presence of a jammer and tested the effectiveness of their approach in realistic scenarios under hardware constraints \cite{Chen_Li_Xu_Xiao_2018}.

\subsection{Motivation and Contributions}

Liu \textit{et al.} explored a power control strategy to mitigate jamming attacks in CR networks \cite{s23020807}. However, the proposed cooperative mechanism may not suit the dynamic nature of CIoT networks, where users frequently join and leave on an ad-hoc basis, and assumes a unified objective among SUs which may not hold in practice. To date, only a handful of studies, namely \cite{Ali_Naser_Muhaidat_2023,Chen_Li_Xu_Xiao_2018}, have directly investigated anti-jamming approaches specifically tailored for CIoT networks, pointing to a significant research gap in this area. Ali \textit{et al.} introduced a DRL algorithm with an emphasis on energy efficiency, aiming to implement a system architecture that conserves energy \cite{Ali_Naser_Muhaidat_2023}. Nonetheless, we believe that focusing solely on energy-efficient DRL algorithms may not adequately account for the broader spectrum of energy constraints, particularly in scenarios involving battery-operated CIoT devices. Furthermore, the strategy in \cite{Chen_Li_Xu_Xiao_2018} that increases CIoT device transmit power to counteract jammers, presents substantial challenges in energy-limited settings. Additionally, such an approach to power control is particularly challenging in underlay CR environments, where SUs and PUs share the spectrum. Studies such as \cite{Ali_Naser_Muhaidat_2023, Chen_Li_Xu_Xiao_2018, 9558789} have overlooked spectrum-sharing scenarios in CR networks. Moreover, \cite{Chen_Li_Xu_Xiao_2018, 9558789} overlook the impact of channel fading effects, further underscoring the need for a comprehensive solution.

To the best of the authors' knowledge, no previous research has considered designing an intelligent algorithm that aims to maximize the CIoT network's throughput under interference limits, energy constraints, and jamming attacks. The main contributions of the paper are summarized as follows:

\begin{itemize}
    \item We introduce a novel strategy for a battery-powered CIoT transmitter, enabling autonomous decision-making to enhance long-term network throughput within spectrum-sharing limits, mitigate jamming interference, and extend network life. Our method uniquely positions the CIoT transmitter to counter jamming attacks directly in the same channel. Additionally, we evaluate the influence of small-scale fading and implement an effective EH model, allowing the CIoT transmitter to exclusively harvest energy from active RF transmissions without dedicating infrastructure for charging.
    \item We formulate the throughput optimization problem for the CIoT transmitter while taking into account factors such as channel occupancy, jamming attacks, channel gain, energy arrival, battery limits, and interference constraints. This approach directs the power control and transmission decisions in the CIoT network, modelled as a model-free Markov decision process (MDP).
    \item We develop a novel DRL algorithm to learn the optimal transmission strategy that maximizes throughput without prior knowledge about the channel or jamming patterns. Our algorithm uses a DDQN designed to enable faster convergence and enhance the algorithm's energy efficiency. Additionally, we introduce an innovative upper confidence bound (UCB) strategy, named UCB interference-aware (UCB-IA), meticulously designed to efficiently mitigate jamming interference and optimize the decision-making framework within the CIoT environment. 
     \item We offer an analysis of convergence and performance of the proposed DRL algorithm that is benchmarked against alternative methodologies found in existing literature across various test scenarios. For performance evaluation, we utilize metrics such as average sum rate, average achievable reward, and jammer interference ratio. Simulations show that under the presence of jamming attacks, the proposed learning algorithm can dynamically choose between data transmission and EH and perform power control to find the optimal solution for the network.
\end{itemize}

The rest of the paper is organized as follows. Section II provides the related work. Section III describes the system model, Section IV showcases the optimization problem formulation, Section V presents the proposed DRL-driven throughput optimization under malicious jamming attacks, Section VI presents the simulation model and comprehensive analysis of the paper's results, and finally, we conclude the paper in Section VII. To enhance clarity, the notations used in this paper are clearly outlined in Table~\ref{tab:notations}.

\begin{table}[t!]
\caption{Notations}
\label{tab:notations}
\resizebox{\columnwidth}{!}{
\begin{tabular}{l|l}
\hline
\textbf{Notation}      & \textbf{Definition}                              \\ \hline
$T$                    & Number of time slots                             \\
$t$                    & Time slot number                                 \\
$\tau$                 & Duration of each time slot                       \\
$B_{max}$              & Battery capacity of CIoT Tx                       \\
$B_{t}$                & Battery level at time slot $t$                       \\
$L$                    & Number of PU transmission slots                  \\
$\zeta$                & Number of slots the jammer attacks \\
$\zeta_{\text{max}}$   & Maximum number of slots that jammer targets \\
$P_p^t$                  & PU transmit power in time slot $t$                               \\
$P_s^t$                  & CIoT Tx transmit power in time slot $t$                            \\
$P_j^t$                & Transmit power of the jammer in time slot $t$                 \\
$\omega_p^t$           & PU status indicator in time slot $t$             \\
$\omega_j^t$           & Jammer status indicator in time slot $t$             \\
$I_{th}$               & Interference threshold                           \\
$d_t$                  & CIoT device's decision to transmit/harvest               \\
$R^t$                  & Achievable rate of the CIoT Tx in time slot $t$\\
$k$                    & Number of slots used by the CIoT Tx                \\
$g_{sp}^t$             & Channel power gain between the CIoT Tx and PU Rx   \\
$g_{ss}^t$             & Channel power gain for the CIoT Tx-Rx pair         \\
$g_{ps}^t$             & Channel power gain for the PU Tx and CIoT Rx       \\
$\sigma^2$             & Channel noise variance                           \\
$e_t$                  & Energy harvested in time slot $t$                \\
$E_{max}$              & Maximum possible energy to be harvested          \\

$\boldsymbol{\mathcal{S}}$       & State space                                      \\
$\boldsymbol{\mathcal{A}}$       & Action space                                     \\
$\boldsymbol{\mathcal{P}}$          & Set of state transition probabilities            \\
$\boldsymbol{\mathcal{R}}$       & Set of possible rewards                          \\
$s_t$                  & State of the environment at time slot $t$        \\
$a_t$                  & Action taken by the CIoT agent at time slot $t$ \\
$r_t$                  & CIoT agent's reward at time slot $t$            \\

$\gamma$               & Discount factor                                  \\
$\boldsymbol{\theta}$  & Parameters of DDQN                                \\
$\boldsymbol{\theta}'$ & Parameters of Target DDQN                         \\
$\alpha$               & Leakiness parameter                              \\
$\kappa$               & Update rate of Target DDQN                        \\
$\mathcal{L}$          & Training loss function                           \\
$\mathcal{M}$          & Experience replay memory                         \\
$m$                    & Memory size                                      \\
$\eta$                 & Learning rate                                    \\
$\hat{r}_a$         &  Empirical mean reward of action $a$
                  \\
$C_a$         &  Number of times action $a$ is chosen
                  \\
$U_a$         &  Actual expected reward of action $a$
                  \\
$\hat{\lambda}_a$         &  Empirical mean jammer interference of action $a$
                  \\
$\overline{U}_a$         &  Expected jammer interference of action $a$
                  \\
$c'$             & UCB adjustment term                                \\ \hline
\end{tabular}
}
\end{table}

\section{System Model}

\subsection{Cognitive IoT Network}

\begin{figure*}
    \centering
    \includegraphics[scale=0.95]{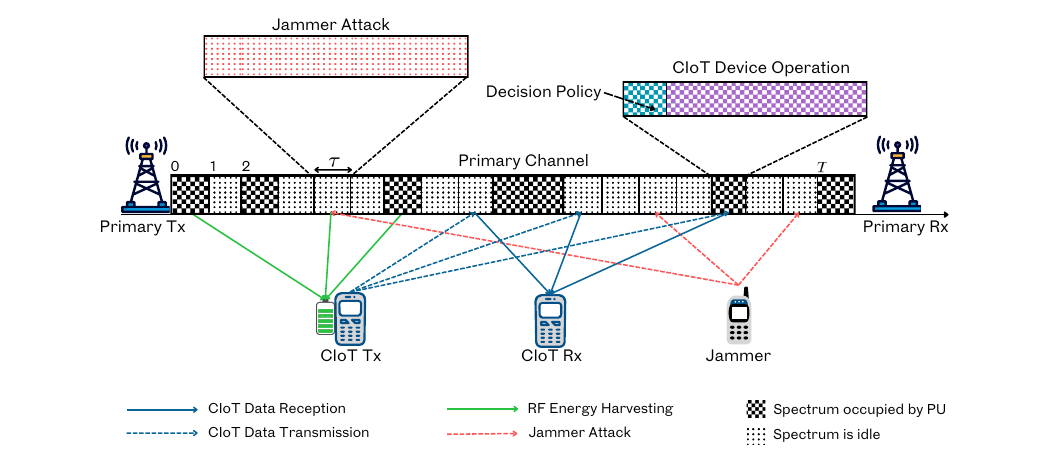}
    \caption{System Model}
    \label{fig:sys_mod}
\end{figure*}

Consider the underlay CIoT network depicted in Fig.~\ref{fig:sys_mod} with a CIoT transmitter-receiver (Tx-Rx) pair, where the Tx is equipped with a rechargeable battery and is capable of RF energy harvesting (EH). The CIoT network shares the dedicated spectrum with the primary network, which consists of a primary Tx-Rx pair. The CIoT Tx possesses the capability to autonomously and dynamically regulate its transmit power $P_s^t$. The CIoT system operates in a time-slotted manner, with $T$ time slots each of duration $\tau$ seconds. For the CIoT Tx, every time slot $t$ encompasses two distinct phases: the decision-making phase and the operation phase. In the decision-making phase, the CIoT device chooses either to transmit data or to harvest energy, following a specific policy. The decision made by the CIoT Tx $d_t$ is set to 0 when transmitting messages and to 1 when harvesting energy. Consequently,
\begin{equation}
  d_t =
  \begin{cases}
    0 & \text{Data transmission mode in time slot $t$}, \\
    1 & \text{Energy harvesting mode in time slot $t$.}
  \end{cases}
\end{equation}
The operation phase involves the CIoT device carrying out the chosen decision $d_t$. We assume the number of slots used by the PU Tx is $L$ where $1 < L < T$ and the PU Tx can consistently transmit at a power level of $P_p^t$ at any slot. 

In underlay CR, the CIoT Tx can use the same time slot as the PU Tx as long as it adheres to the interference threshold $I_{th}$. Therefore, the CIoT device needs to decide on its transmission power $P_s^t$ such that it adheres to the interference constraint given as
\begin{equation}
    P_{s}^tg_{sp}^t \leq I_{th},
\end{equation} 
where $g_{sp}^t$ is the channel power gain between the CIoT Tx and PU Rx. The PU status indicator is defined as
\begin{equation}\label{eq:PU_indicator}
  \omega_p^t =
  \begin{cases}
    1 & \text{if the PU Tx is using time slot $t$}, \\
    0 & \text{otherwise.}
  \end{cases}
\end{equation}
The power gains of the channel between the CIoT Tx-Rx pair is $g_{ss}^t$, the PU Tx and the CIoT Rx is $g_{ps}^t$, and the CIoT Tx and the PU Rx is $g_{sp}^t$, are characterized as Rayleigh fading channels that are independently and identically distributed (i.i.d). It is assumed that these channel power gains remain constant within a time slot, but may vary independently of one time slot to another. 

\subsection{Jamming Model}

We consider a jammer that launches jamming attacks on the CIoT network, as shown in Fig.~\ref{fig:sys_mod}. The goal of the jammer is to make the shared spectrum ``appear" busy and prevent CIoT Tx from accessing it and potentially depleting the CIoT device's battery. We consider that the jammer can only target the CIoT transmissions. This could be attributed to the potential severe penalties faced by attackers upon identification by PUs, or the inability to approach PUs. Moreover, in practice, methods such as cyclostationary detection or matched filter detection can be used by the jammers to differentiate between the PUs and CIoT transmissions. Therefore, at the beginning of each time slot, the jammer determines whether the PU is active or not before launching the jamming attack. Should a jammer initiate an attack, it will attack for the duration of the entire time slot \cite{9099069}. The jammer launches a jamming attack with power $P_j^t$. The CIoT Tx has no prior information about which time slots will be subjected to jamming attacks.

Attackers tend to favor a random jamming approach, as it enables intermittent periods of inactivity \cite{9099069}. This strategy not only prolongs the jammer's lifetime but also decreases the chance of its detection. Therefore, we consider the scenario of intermittent jamming, where the jammer randomly toggles between periods of uniformly distributed jamming and rest. That is, the jammer engages in attacks for a duration of $\zeta \sim U (0, \zeta_{\text{max}})$ slots, followed by a period of rest lasting $T-\zeta$ slots, where $\zeta_{\text{max}}$ denotes the maximum number of time slots that the jammer is capable of sustaining jamming attacks. Therefore, the probability of launching an attack is $U(0, \zeta_{\text{max}})/T$. By incorporating randomness into the jammer's actions, the jamming behavior becomes less predictable for the CIoT Tx. In this paper, the impact of one jammer launching attacks is analogous to multiple coordinated jammers targeting specific time slots for their attacks. This allows our system to be scalable to incorporate several coordinated jammers, with each one executing a jamming attack in individual time slots. 

The CIoT agent determines the jammer's status indicator at the beginning of each time slot using broadband sensing capabilities \cite{9099069}. The jammer's status indicator, which reflects whether the jammer is launching a jamming attack in a time slot $t$ is defined as follows
\begin{equation}\label{eq:jam_indicator}
  \omega_j^t =
  \begin{cases}
    0 & \text{Jammer is launching an attack at time slot $t$}, \\
    1 & \text{otherwise,}
  \end{cases}
\end{equation}
where $\omega_j^t=0$ indicates that the CIoT agent detects the presence of a jamming attack, following the probability $U(0, \zeta_{\text{max}})/T$, and $\omega_j^t=1$ indicates that the CIoT agent does not detect a jamming attack, with a probability of $1-U(0, \zeta_{\text{max}})/T$.

\subsection{Energy Harvesting Model}
The CIoT transmitter is capable of charging its finite battery, denoted by $B_{max}$, through RF EH \footnote{The implementation of circuits responsible for the process of RF EH is beyond the scope of this paper.}
Initially, the harvested energy is set to zero. It is assumed that the amount of energy collected in each time slot, $e_t$, follows a uniform distribution ranging from $0$ to $E_{max}$, where $e_t \sim U(0, E_{max})$. However, the maximum possible energy to be harvested $E_{max}$ is affected by the RF signals, which are influenced by the activity of both the jammer and the PU Tx within each time slot. Thus, the value of $E_{max}$ varies depending on the following
\begin{equation} 
E_{max} =
\begin{cases}
    P_p^t & \text{if}~ \omega_p^t=1\text{, PU transmitting}, \\
    P_j^t & \text{if}~ \omega_j^t =0\text{, jammer is attacking},\\
    P_p^t +  P_j^t & \text{if}~ \omega_p^t=1 ~\text{and}~ \omega_j^t = 0.
\end{cases}
\end{equation}
The energy collected at any given time slot is kept in the battery and is solely used for operation in future time slots. However, due to constraints in the hardware, the CIoT Tx is unable to perform RF EH and access the spectrum opportunistically at the same time. It is important to highlight that the energy harvesting process does not lead to any additional energy consumption for PUs or other devices \cite{9606870}.

The initial battery level of the CIoT Tx is denoted as $B_0$, with $B_t$ representing the energy available in the battery at the $t$-th time slot. Following the approach in \cite{chu_eh_ciot_ma_2019}, we assume the battery to be ideal, meaning there are no losses of energy during its storage or retrieval. For the CIoT Tx, energy consumption is exclusively due to data transmission. Moreover, any harvested energy that exceeds the battery's maximum capacity is deemed to be discarded. Additionally, the concept of normalized time slots is used, which permits treating the terms ``energy" and ``power" as synonymous \cite{chu_eh_ciot_ma_2019}. At any given time slot $t$, the selected transmission power $P_s^t$ by the CIoT Tx must not exceed the battery's total energy, $B_t$. That is,
\begin{equation}
    0\leq (1-d_t) P_s^t \tau \leq B_t,
\end{equation}
where $\tau$ represents the duration of each time slot. The increase or decrease in the battery's energy level is determined by the CIoT device's decision $d_t$ to harvest energy/transmit data at a time slot $t$. In the subsequent time slot, $t+1$, the amount of energy available is updated based on $d_t$. Therefore, the battery update is as follows
\begin{equation} \label{eq:eq01} 
    B_{t+1} = \text{min}\big\{
    B_t + d_t e_t -(1-d_t)P^t_s\tau, B_{max}\big\}.
\end{equation}
Consequently, the total energy consumed by the CIoT device cannot exceed the total energy collected in the battery. That is,
\begin{equation}
    \sum_{t=1}^k P_s^t\tau \leq B_0 + \sum_{t=0}^{k-1} e_t, \forall k.
\end{equation}
where $k$ represents the total number of time slots in which the CIoT device decides to transmit.
\subsection{Transmission Model}
The CIoT Tx is tasked with setting its transmit power $P_s^t$ such that it maximizes its sum rate under jamming attacks and avoids interfering with the primary network. During an idle $t$-th time slot, the achievable sum rate by the CIoT Tx is
\begin{equation} 
 R_0^t = \log_2\bigg(1+\frac{P_s^t g_{ss}^t}{\sigma^2}\bigg),
\end{equation} 
where $\sigma^2$ denotes the variance of the channel noise. Should the PU Tx occupy the channel during the $t$-th time slot, the CIoT Tx's achievable sum rate will be reduced due to PU interference, as described by
\begin{equation} 
 R_1^t = \log_2\bigg(1+\frac{P_s^t g_{ss}^t}{P^t_p g_{ps}^t + \sigma^2}\bigg).
\end{equation}
Hence, the achievable sum rate for the CIoT Tx during a time slot $t$ can be expressed as
\begin{equation}
    R^t = \omega_j^t(1-d_t)\big{[}(1-\omega_{p}^t)R^t_0+\omega_{p}^tR^t_1\big{]}.
    \label{eq:tot_rate}
\end{equation}
According to (\ref{eq:tot_rate}), if the jammer is launching an attack at time slot $t$, i.e., $\omega_j^t=0$, then the achievable sum rate $R_t$ of the CIoT Tx effectively drops to zero.

\section{Optimization Problem Formulation}
In this section, we aim to define the optimization problem, to develop a dynamic algorithm that maximizes the total sum rate of the CIoT network. This algorithm will address challenges such as jamming attacks, channel fading, interference, and energy constraints. The CIoT device needs to learn how to optimize both its transmission power \(P_s^t\) and decision \(d_t\) to maximize the throughput of the network. That is, the CIoT device must strategically choose between transmitting data, which consumes battery and may be affected by jamming, or harvesting energy, sacrificing immediate transmission opportunities. Therefore, the optimization problem can be formulated as follows

\begin{subequations} 
\label{eqn:optim}
\begin{align}
   &\max_{P^t_s} \sum_{t=1}^{T} \omega_j^t(1-d_t)\big{[}(1-\omega_{p}^t)R^t_0+\omega_{p}^tR^t_1\big{]}, \label{maximization}\\
   &\text{s.t.  } \sum_{t=1}^{k} P^t_s\tau \leq B_0+\sum_{t=0}^{k-1}e_t, ~~\forall k,  \label{constraint1}\\
   & ~~~~~0\leq (1-d_t)P^t_s\tau\leq B_t,~~ d_t\in I\triangleq\{0,1\}, \label{constraint2}\\
   & ~~~~~d_t = 1 , ~~ \forall \omega_j^t=0, \label{constraint3}\\
   & ~~~~~\omega_{p}^tg_{sp}^tP^t_s\leq I^t_{th},~~ \omega_{p}^t\in\Omega\triangleq\{0,1\} .\label{constraint4}
\end{align}
\end{subequations}

Constraint (\ref{constraint1}) ensures that the transmission power $P^t_s$ of the CIoT device remains within the bounds set by the initial battery level $B_0$ and the energy harvested over all the time slots within the period. Constraint (\ref{constraint2}) requires that the CIoT device's transmission power does not exceed the current battery level $B_t$ in any specific time slot $t$. Constraint (\ref{constraint3}) protects against the use of time slots when the jammer is active. Moreover, (\ref{constraint4}) dictates that the CIoT Tx complies with the interference threshold $I_{th}$ to avoid causing interference to the PU during each time slot $t$. In the subsequent discussion, we introduce a solution for the optimization problem presented in (\ref{maximization}) by employing a model-free Markov Decision Process (MDP).

\subsection{The Model-Free Markov Decision Process (MDP)}
 The process of learning the optimal strategy that maximizes the sum rate of the CIoT network can be formulated as an MDP. The MDP is represented by the tuple $(\boldsymbol{\mathcal{S}}, \boldsymbol{\mathcal{A}}, \boldsymbol{\mathcal{P}}, \boldsymbol{\mathcal{R}}, T)$, where $\boldsymbol{\mathcal{S}}$ signifies the set of CIoT environment states, $\boldsymbol{\mathcal{A}}$ represents the set of possible actions by the CIoT agent \footnote{We use CIoT agent/device to refer to the transmitter within the network under study.}, $\boldsymbol{\mathcal{P}}$ represents the set of state transition probabilities, $\boldsymbol{\mathcal{R}}$ includes the rewards for specific state-action combinations, and $T$ indicates the time step. In practical CR contexts, it is notably difficult to accurately determine the probability density functions of energy and channel fading \cite{eh_drl_sim_ppr}. Moreover, under jamming attacks, calculating the transition probabilities necessitates having full and accurate knowledge of the jamming pattern. Given that the jammer aims to disrupt the transmissions of the CIoT network and will therefore not reveal their information, obtaining the state transition probabilities proves to be impractical. Therefore, as a solution, a model-free MDP approach is utilized, and deep reinforcement learning (DRL) is employed to infer $\boldsymbol{\mathcal{R}}$ based on $\boldsymbol{\mathcal{S}}$ and $\boldsymbol{\mathcal{A}}$, in the absence of $\boldsymbol{\mathcal{P}}$. Consequently, the CIoT device is designed to deduce a policy $\pi: \boldsymbol{\mathcal{S}} \to \boldsymbol{\mathcal{A}}$ through ongoing interactions with the environment and learning from them the actions that yield the most cumulative reward. This leads to a revised model-free MDP: $(\boldsymbol{\mathcal{S}}, \boldsymbol{\mathcal{A}}, \boldsymbol{\mathcal{R}}, T)$. The elements of this MDP tuple are detailed below.

\textbf{State Space $\boldsymbol{\mathcal{S}}$:} Within each time slot, the CIoT device, acting as a learning entity, assesses the state of the unknown environment (channel) for its decision-making process. The state space comprises all possible states across the $T$ time slots. For any given environment state $s_t$, the CIoT agent must consider several factors: the current level of battery $B_t$, the energy harvested in the preceding time slot $e_{t-1}$, the presence of a PU Tx, the presence of jamming attacks, and the channel power gains denoted as $g_{ps}^t, g_{sp}^t, g_{ss}^t$. Hence, the CIoT environment's state at the $t$-th time slot encompasses these elements as follows
\begin{equation}
        s_t=\{ B_t, e_{t-1}, \omega_{p}^t, \omega_{j}^t, g_{ps}^t, g_{sp}^t, g_{ss}^t\}.
\end{equation}

\textbf{Action Space $\boldsymbol{\mathcal{A}}$:} The action space encompasses every possible action the CIoT agent is capable of performing. For optimizing throughput, it is appropriate to consider both the decision $d_t$ and the transmit power $P_s^t$ as integral elements of the action. Based on the environment state $s_t$, the CIoT agent must choose either to transmit data ($d_t=0$) or to harvest energy ($d_t=1$), and set its transmission power $P_s^t$ appropriately. Thus, the CIoT agent's action at time slot $t$ is defined as
\begin{equation}
    a_t = [d_t, P_s^t],~ \text{where}~ d_t\in I\triangleq\{0,1\}, ~\text{and}~ P^t_s \in P.
\end{equation}
$P$ is the set of possible transmission powers by the CIoT agent.

\textbf{Reward $\boldsymbol{\mathcal{R}}$:} The CIoT agent evaluates the effectiveness of its chosen actions through the rewards received, using this feedback to refine its decision-making strategy. As such, the reward is defined by the data transmission rate achieved by the CIoT Tx, under the condition that it adheres to the constraints specified in (\ref{eqn:optim}). Choosing to harvest energy results in a reward of 0. Should the CIoT Tx engage in an action $a_t$ that breaches the constraints outlined in (\ref{eqn:optim}), it incurs a negative reward as a form of penalty. Thus, the reward $r_t$ for the CIoT agent in each time slot $t$ is determined as follows 

\small\begin{equation} \label{eq:reward}
 r_t = \begin{cases} R_0^t & d_t=0,\omega_{p}^t=0,\omega_{j}^t=1,0\leq P_s^t \tau \leq B_t,\\
 R_1^t & d_t=0,\omega_{p}^t=1,\omega_{j}^t=1,0\leq P_s^t\tau\leq B_t, P_s^t g_{sp}^t\leq I_{th},\\
 0 & d_t=1,P_s^t\tau>B_t,\\
 -\phi & \text{others}.
\end{cases}
\end{equation}\normalsize

\textbf{Time Step $T$:} The transition from a current time slot $t$ to the following slot $t+1$ constitutes a discrete time step. In this framework, we analyze all possible state-action pairs within the duration of $T$ time slots, iterating through these pairs systematically for each time slot increment.

\section{DRL-Driven Throughput Optimization Under Malicious Jamming}

In the specified model-free MDP, the CIoT agent faces the challenge of assessing the value of state-action pairs without prior knowledge of $\boldsymbol{\mathcal{P}}$. However, through reinforcement learning (RL), the CIoT agent can approximate the state-value function and learn a policy $\pi$ that guides action selection $a_t$ per the current CIoT environment state $s_t$. The aim is to maximize the reward (sum rate) given by utilizing a policy $\pi$. This involves navigating challenges like malicious jamming while accommodating considerations for spectrum sharing, energy limitations, the variability of energy arrival, and channel fading conditions.

A policy $\pi$ is a function that assigns probabilities to taking action \(a\) when in state $s$, mapping each state-action pair to the probability $ \pi (s,a) $ of selecting action $(a \in \boldsymbol{\mathcal{A}})$ in state \(s \in \boldsymbol{\mathcal{S}}\). For the model-free MDP, we can define the expected value of the state-action value function, also known as the Q-function or Bellman equation, as
\begin{equation} 
Q^\pi (s,a) = \mathbb{E}[r_t + \gamma Q^\pi(s_{t+1}, a_{t+1})|s_t=s,a_t=a],
    \label{eqn:q_learning}
\end{equation}
where $r_t$ represents the immediate reward for taking action $a_t$ in state $s_t$. The term $\gamma Q^\pi(s_{t+1}, a_{t+1})$ reflects the discounted expectation of future rewards, with $\gamma$, a discount factor ranging between 0 and 1, indicating the relative significance of future rewards compared to immediate ones. Higher $\gamma$ values prioritize long-term rewards more heavily. The objective for the CIoT agent is to identify the optimal action $a_t$ that maximizes the Q-value at each time slot $t$, meaning
\begin{equation}
    a^* =\arg \underset{a \in \boldsymbol{\mathcal{A}}}{\max} \big\{Q^\pi (s,a)\big\}.
\end{equation}

Through the Q-learning algorithm, the CIoT agent determines the Q-value at every step and records it in a Q-table to identify the best solution. The most basic method for updating the action-value function is described as \cite{Sutton_Barto_2020}
\begin{equation}
\begin{split}
 &Q^\pi (s,a) =  \\
 Q^\pi &(s,a) + \eta \bigg [ r_{t+1} + \gamma \max_{a} Q^\pi(s_{t+1}, a) - Q^\pi (s,a) \bigg],
 \end{split}
    \label{eq:ql_update}
\end{equation}
where $\eta \in [0,1]$ represents the learning rate. However, Q-learning may encounter slow convergence rates in finding the optimal actions for problem resolution \cite{9558789}. Therefore, we explore deep Q-learning, integrating principles from both RL and deep learning, to estimate the Q-value function through the use of a deep neural network, also known as a double deep Q-network (DDQN). This approach aims to enhance the approximation of the Q-value function for more efficient training.

\begin{figure*}
    \centering
    \includegraphics[width=2\columnwidth]{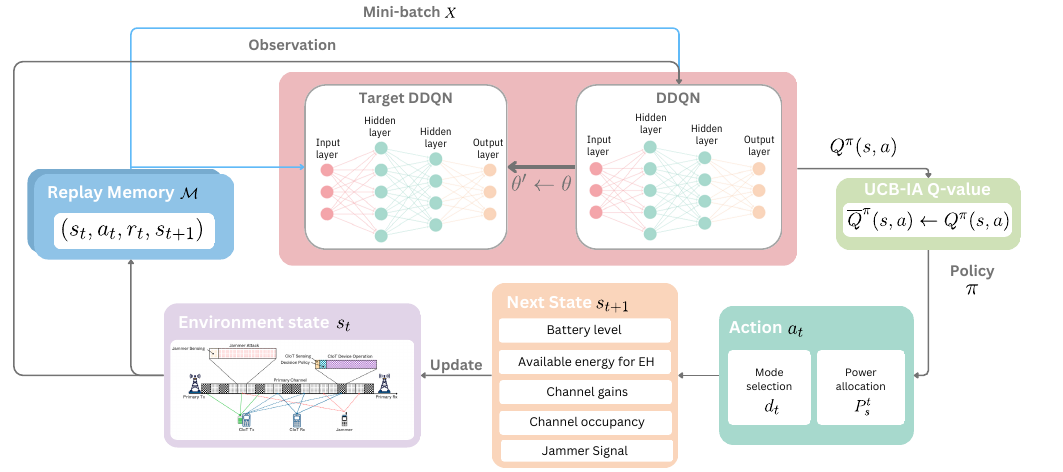}
    \caption{The proposed DRL algorithm featuring the UCB-IA action exploration strategy.}
    \label{fig:dqn_model}
\end{figure*}

Fig.~\ref{fig:dqn_model} provides a detailed view of the employed DRL algorithm, including our proposed UCB-IA exploration strategy, which will be discussed in the next subsections.

\subsection{The proposed DDQN-Driven DRL Approach}
In this subsection, we present our DDQN architecture, designed to determine the best policy for enhancing transmission efficiency within the CIoT network under malicious jamming attacks. Our DDQN aims to predict the total expected reward (Q-value) for every possible action $a_t$ within a specific state $s_t$. Essentially, this involves adjusting the DDQN's parameters $\boldsymbol{\theta}$ iteratively such that
\begin{equation}
Q^\pi(s,a;\boldsymbol{\theta}) \approx Q^\pi(s,a).
\end{equation} The parameters of the DDQN, $\boldsymbol{\theta} = \{ \textbf{W}^{(i)}, \textbf{b}^{(i)} \}$, for the layers $i = \{1, \ldots, 4\}$, signify the network's weights and biases. The DDQN employs a fully connected neural network architecture. Its input layer features $j$ neurons that match the state space $\boldsymbol{\mathcal{S}}$ dimensionality. Additionally, the neural network includes two hidden layers with $h_1$ and $h_2$ neurons, respectively. The output layer consists of $z$ neurons.

To promote faster convergence of the DDQN and enhance the stability of the training process, we utilize a weight initialization method called Kaiming (He) initialization \cite{He2015}. This technique initializes the DDQN's weights by drawing from a Gaussian distribution $\mathcal{N}(0,\frac{2}{\nu_i})$, where $\nu_i$ denotes the count of input neurons for each layer $i$. To introduce non-linearity into the neural network architecture, we consider a rectified linear unit (ReLU) function. The basic ReLU activation function is expressed as $f(x) = \max(0,x)$, offering computational advantages over other activation functions such as sigmoid or hyperbolic tangent (tanh). This is because it implements a simple threshold at zero, significantly accelerating the training process. However, the ReLU function faces the issue of "dying ReLU" (neurons that cease to activate). To overcome this, we employ a leaky ReLU function to ensure active participation of all neurons in the learning process, which in turn facilitates quicker convergence and enhances the learning dynamics. The leaky ReLU activation function is defined as
\begin{equation} 
f(x) = \begin{cases}
    x, & \text{if } x \geq 0, \\
    \alpha x, & \text{if } x < 0,
\end{cases}
\label{eqn:leaky_relu}    
\end{equation}
where $\alpha \in \{0,1\}$ represents the "slope" parameter that controls the degree of "leakiness" in the ReLU function.

To enhance the training process of the CIoT agent and ensure data efficiency, we utilize empirical replay. This involves using a memory buffer $\mathcal{M}$ of capacity $m$ to store previous experiences in the form of tuples $(s_t, a_t, r_t, s_{t+1})$. Once the memory reaches capacity, mini-batches of experiences are randomly sampled from the data set of state-action pairs and employed in the training process to update the DDQN's parameters. By adopting this method, we can effectively eliminate temporal correlations in the training data, significantly lowering the risk of instability during the training phase \cite{chen_power_control}.

Deep reinforcement learning (DRL) is known to be unstable or even diverge since it uses a non-linear deep neural network for Q-function approximation. Several factors contribute to this instability. Minor adjustments to the Q-function can drastically alter the policy, affecting data distribution. Additionally, interdependence between action-values and target values derived from maximizing $Q^\pi$ over all possible actions in the next state worsens instability. Therefore, to overcome this challenge, we utilize a Target DDQN during training. The Target DDQN is used to compute the target optimal Q-function as 
\begin{equation}\label{eq:targetQ}
    Y = r_t + \gamma \max_{\textbf{a}\in \boldsymbol{\mathcal{A}}}~ Q^\pi(\textbf{s}_{t+1},\textbf{a}_{t+1};\boldsymbol{\theta}'),
\end{equation}
where $\boldsymbol{\theta}'$ are the Target DDQN's parameters. At the beginning of training, the Target DDQN is a mirror of the DDQN, i.e., $\boldsymbol{\theta}'= \boldsymbol{\theta}$. With the progression of training, the Target DDQN's parameters are updated at a slower rate than those of the DDQN, covering several training steps. The frequency of updates to the Target DDQN is denoted by $\kappa$.

To update the DDQN's parameters, we utilize the mean squared error (MSE) loss $\mathcal{L}(.)$ during training to quantify the deviation between the estimated Q-values and the target Q-values across a mini-batch of state-action pairs $(\textbf{s},\textbf{a})$.
\small \begin{equation}   
    \mathcal{L}(\boldsymbol{\theta}) = \mathbb{E}\Bigg[ \bigg[ Y -Q^\pi(\textbf{s},\textbf{a};\boldsymbol{\theta})\bigg]^2\Bigg].
    \label{eqn:loss_fn}
\end{equation}\normalsize
During the training phase, the objective is to minimize the loss described in (\ref{eqn:loss_fn}) across a mini-batch of state-action pairs. This entails \begin{equation}
    \boldsymbol{\hat{\theta}}= \arg \underset{\boldsymbol{\theta}}{\min}~ \mathcal{L}(\boldsymbol{\theta};\textbf{s},\textbf{a}).
\end{equation} The backpropagation algorithm \cite{bishop2006pattern} is effective in computing $\nabla_{\boldsymbol{\theta}}\mathcal{L}(\boldsymbol{\theta};\textbf{s},\textbf{a})$, the gradient of the loss with respect to the DDQN's parameters given the state-action pairs. Once $\nabla_{\boldsymbol{\theta}}\mathcal{L}(\boldsymbol{\theta};\textbf{s},\textbf{a})$ is obtained, stochastic gradient descent (SGD) \cite{eh_drl_sim_ppr} can then be employed to update the DDQN's parameters accordingly as
\begin{equation} 
\begin{split}
    \boldsymbol{\theta} = \boldsymbol{\theta} -& \eta\nabla_{\boldsymbol{\theta}}\mathcal{L}(\boldsymbol{\theta};\textbf{s},\textbf{a}), \\
    \text{where}~\boldsymbol{\theta} &= \{ \textbf{W}^{(i)}, \textbf{b}^{(i)} \}, \\ \text{for}~i &= \{1, \ldots, 4\}.
    \label{eqn:sgd}
    \end{split}
\end{equation}
$\eta \in \{0,1\}$ is the learning rate that determines the update rate in each iteration of SGD. 

In this paper, we utilize an advanced SGD-based parameter update technique known as adaptive moment estimation (Adam) due to its faster computation time \cite{10279156,10437464}. Additionally, we incorporate a carefully designed learning rate scheduler. Initially, it sets a learning rate to ensure stable learning in the early phases. As training progresses, the scheduler dynamically adjusts the learning rate based on factors such as model performance and a specified patience period. This approach facilitates efficient convergence and enhances overall performance. The block diagram of the DDQN approach, thoroughly explained in this subsection, is presented in Fig. \ref{fig:dqn_model}.

\subsection{UCB-IA: Interference-Aware Action Exploration Strategy}
As it is well known in the literature, there is a trade-off between the exploration of new actions in the action space (i.e. learning their mean reward) and the exploitation of known actions (i.e. achieving the highest empirical rewards). If the expected rewards of actions were known, the optimal policy would always choose the action that offers the highest expected reward. For the CIoT agent to explore the environment, discover optimal strategies, and balance the exploitation-exploration tradeoff we employ the principles of the upper confidence bound (UCB) algorithm. 

The classical UCB algorithm adjusts the Q-values based on $\overline{Q}^\pi (s,a) = Q^\pi (s,a) + U_a^t$,
where $U_a^t$ is the actual expected reward calculated as $U_a^t = \hat{r}_a^t+\sqrt{\frac{c'\ln{t}}{C_a^t}}$,
where $c'$ is a hyperparameter of the UCB algorithm. The actual expected reward $U_a^t$ is a combination of the expected reward $\hat{r}_a^t$ and an adjustment term which depends on the time period number, i.e., frame number * $T$ + $t$, and the number of times action $a$ has been chosen $C_a^t$. If action $a_t$ has been chosen $C_a^t$ times by the end of time slot $t$ (i.e., from 0 to $t$), then $\hat{r}_a^t = (\sum_{i=1}^{C_a^t}r_{a,i}^t)/C_a^t$,
where $r_{a,i}^t$ is the reward of action $a_t$ in the $i$-th time it was chosen. Then, the action returned by the UCB algorithm is used in the DDQN training and the action count $C_a^t$ and expected reward $\hat{r}_a^t$ are updated accordingly.

In this work, we propose a novel variant of the UCB algorithm called Interference-Aware UCB (UCB-IA), which is illustrated in Algorithm~\ref{alg:algorithm1}. The proposed UCB-IA exploration-exploitation strategy not only takes the expected reward $\hat{r}_a^t$ to update the Q-value, but also takes the actual expected jammer interference $\hat{\lambda}_a^t$. This addition allows the agent to refine its performance by identifying actions that experience jammer interference in any state $s_t$, thereby adjusting the Q-values to achieve a higher reward rate while minimizing jammer interference. Therefore, under the UCB-IA strategy, the actual expected reward $\overline{U}_a^t$ is defined as 
\begin{equation}
    \overline{U}_a^t = \hat{r}_a^t\hat{\lambda}_a^t+\sqrt{\frac{c'\ln{t}}{C_a^t}}.
\end{equation} We express the actual expected jammer interference $\hat{\lambda}_a^t$ as
\begin{equation}
    \hat{\lambda}_a^t = \frac{\sum_{i=1}^{C_a^t}\lambda_{a,i}^t}{C_a^t},
\end{equation} where $\sum_{i=1}^{C_a^t}\lambda_{a,i}^t$ represents the number of times the CIoT agent has experienced a jamming attack as a result of action $a$. Therefore, $\hat{\lambda}_a^t$ is a value between 0 and 1. Thus, for our proposed UCB-IA algorithm the adjusted Q-value is given by
\begin{equation}
    \overline{Q}^\pi (s,a) = Q^\pi (s,a) + \overline{U}_a^t.
\end{equation}

The adjustment of the Q-value using UCB-IA is illustrated in Fig. \ref{fig:dqn_model}. This modification is applied to the Q-value output of the DDQN architecture, refining the action selection policy in accordance with the proposed UCB-IA strategy.

\begin{algorithm}[t!]
\caption{The proposed UCB-IA-driven DRL Algorithm (\ref{eqn:optim})}\label{alg:algorithm1}
\textbf{Input:} Cognitive IoT environment simulator and its parameters.

\textbf{Output:} Optimal action $a_t$ in each time slot $t$.

Initialize experience replay memory $\mathcal{M}$ with size $m$.

Initialize battery level with $B_0$

Initialize $\forall \theta \in \boldsymbol{\theta}, \quad \theta \sim \mathcal{N}(0, \frac{2}{v_i})$ and initialize $\boldsymbol{\theta}'$ with $\boldsymbol{\theta}' \leftarrow \boldsymbol{\theta}$.

Initialize $\eta$ and set the scheduler's reduction factor and patience period.

Initialize $\gamma$, $\kappa$, and $c'$.

\For{episode= 1,...,episodes}{
    
    \For{t= 1,...,T}{
        Observe the state $s_t$\;
        \If{$\mathcal{M}$ is not full}{
        
        Sample a random action $a_t$\;
        
        Get the reward $r_t$ using (\ref{eq:reward}) and observe the next state $s_{t+1}$ \;
        
        Store $\mathcal{M}\leftarrow(s_t,a_t,r_t,s_{t+1})$\;}

        \Else{ 
            Calculate $\overline{U}_a^t \leftarrow \hat{r}_a^t\hat{\lambda}_a^t+\sqrt{\frac{c'\ln{t}}{C_a^t}}$;

            Adjust Q-value $\overline{Q}^\pi (s,a) \leftarrow Q^\pi (s,a) + \overline{U}_a^t$;
            
            Get action $a_t$ according to the policy of adjusted Q-value;

            Update action count, $C_a^t \leftarrow C_a^t + 1$;

            Get the reward $r_t$ using (\ref{eq:reward}) and observe the next state $s_{t+1}$ \;
            
            }

            Update $\hat{\lambda}_a^t \leftarrow (\sum_{i=1}^{C_a^t}\lambda_{a,i}^t)/C_a^t$;

            Update $\hat{r}_a^t \leftarrow (\sum_{i=1}^{C_a^t}r_{a,i}^t)/C_a^t$
            
            Sample a mini-batch $X$ from $\mathcal{M}$ \;

            Predict Target Q-values using (\ref{eq:targetQ})\;

            Predict Q-values using $Q^\pi(\boldsymbol{s},\boldsymbol{a};\boldsymbol{\theta})$

            Calculate the loss in (\ref{eqn:loss_fn})\;
            
            Update $\boldsymbol{\theta}$ of DDQN online using (\ref{eqn:sgd})\;

            \If{episode*t \text{mod} $\kappa$ = 0}{
            Update $\boldsymbol{\theta}'$ of Target DDQN online as $\boldsymbol{\theta}' \leftarrow \boldsymbol{\theta}$ \;
             }

}
         Update $\eta$ using scheduler \;

             Update $\epsilon$ \;
            
            Update the state $s_{t+1} = s_t$
}

\end{algorithm}

\section{Simulation Model and Results}
In this section, we assess the effectiveness of our proposed DRL strategy featuring UCB-IA exploration in enhancing the transmission efficiency of the EH-enabled CIoT network detailed in Section II. 
\subsection{Simulation Settings}
The CIoT Tx under study possesses the capability to dynamically select a transmit power $P_s$ from the set $P=\{0.01,0.02,...,0.1\}$ based on the current environment state. Moreover, the CIoT Tx is equipped with a finite-sized battery whose initial level is $B_0 = 0$. We consider this scenario for the starting battery, since it is the worst-case scenario, offering a realistic view of the performance of our proposed DRL approach. The simulation parameters utilized can be found in Table~\ref{tab:simulation_parameters}. The channel power gains $g_{ss}^t$ and $g_{sp}^t$ are distributed exponentially with mean values of 0.1 and 0.2, respectively. For simplicity, we assume $g_{sp}^t= g_{ps}^t$. 

In the DDQN architecture, we employ four fully connected layers, each with a specific number of neurons: $j=7$, $h_1=128$, $h_2=64$, and $z=22$, without loss of generality. The leaky rectified linear unit (ReLU) serves as the non-linear activation function, with a leakiness parameter of $\alpha=0.02$. The mean squared error (MSE) loss function $\mathcal{L}$ in (\ref{eqn:loss_fn}) is employed during training. As for the learning rate, we initialize it with $\eta=4\times 10^{-4}$, followed by a 50\% reduction every 500 episodes. A penalty value of $\phi=7$ is incurred whenever the CIoT agent violates the constraints outlined in (\ref{eqn:optim}) during the training phase. The training process consists of many episodes, each representing a time frame of $T$ time slots. The proposed scheme was implemented on Google Colab using Python version 3.9.3 with a CPU hardware accelerator. Our reinforcement learning environment was developed using the Gymnasium library, and our deep learning models were built with the PyTorch 2.3.0 library.
\begin{table}[t!]
\caption{Simulation Parameters}
\label{tab:simulation_parameters}
\resizebox{\columnwidth}{!}{%
\begin{tabular}{l|l}
\hline
\textbf{Parameters}                       & \textbf{Value}               \\ \hline
Number of time slots $T$                  & 30                           \\
Duration of each time slot $\tau$         & 1 s                          \\
Number of PU transmission slots $L$       & 18                           \\
Transmission power of PU $P_p^t$            & 0.2 W                        \\
Interference threshold $I_{th}$           & 0.01 W                       \\
Initial battery level $B_0$ &  0.0 W   \\ 
Battery capacity $B_{max}$                & 0.5 W                        \\
Transmission power range of CIoT Tx $P_s^t$     & 0.01~$\sim$~0.1 W              \\
Transmission power of jammer $P_j^t$           & 0.1 W                        \\
Maximum time slots under jamming $\zeta_{\text{max}}$ & 12 \\
Noise power $\sigma^2$                    & 1e-3 W                       \\
Experience replay memory size $m$         & 10,000                       \\
Training episodes                         & 2500                         \\
Mini-batch size                           & 200                          \\
Learning rate $\eta$                      & $4* 10^{-4}$                 \\
Learning rate reduction factor            & 50\%                         \\
Learning rate patience period             & 500 episodes                 \\
Penalization $\phi$                       & 7                            \\
Discount factor $\gamma$                  & 0.99                             \\
Leakiness parameter $\alpha$              & 0.02                         \\
Update rate of Target DDQN $\kappa$  & 100                             \\                    
UCB adjustment term $c'$   &  1\\
Channel power gain of CIoT Tx-PU Rx $g^t_{sp}$   &  0.2 W\\
Channel power gain of CIoT Tx-Rx $g^t_{ss}$   &   0.1 W\\
Channel power gain of PU Tx-CIoT Rx $g^t_{ps}$   &  0.2 W\\
Number of neurons & 7, 128, 64, 22
                \\\hline
\end{tabular}
}
\end{table}

In our evaluation, we employ performance metrics such as average sum rate (ASR), average reward, and jammer interference rate across training episodes. The total achievable sum rate of the CIoT Tx agent over $T$ time slots (one episode) is expressed as $\sum_{t=1}^{T} R^t$, where $R^t$ is defined in (\ref{eq:tot_rate}). The ASR is the weighted moving average of the total achievable sum rate. Furthermore, the total reward is calculated as all rewards accumulated by the CIoT Tx agent over one episode, given by $\sum_{t=1}^{T} r_t$, with $r_t$ provided in (\ref{eq:reward}). The average reward is the weighted moving average of the total reward of the CIoT agent. Finally, the jammer interference rate is determined by the number of time slots where the CIoT agent transmits data while the jammer is active, divided by the total number of time slots under jamming attacks. This rate is calculated as $\sum_{t=1}^{T}\lambda^t / \sum_{t=1}^{T}\omega_j^t$, where $\omega_j^t$ is defined in (\ref{eq:jam_indicator}) and $\lambda^t=1$ indicates that the CIoT agent experienced jamming in time slot $t$ (with $\lambda^t=0$ otherwise).

The calculation of the weighted moving average is as follows
\begin{equation}
    \text{average}_{\text{new}} = (1 - \delta) \times \text{average}_{\text{old}} + \delta \times \text{value},
\end{equation}
where $\delta$ represents the significance according to the newest data point, with $1-\delta$ reflecting the importance of the accumulated historical average. In this context, we have chosen $\delta =0.01$. By utilizing the weighted moving average, short-term fluctuations are smoothed out, allowing the underlying trends in both sum rate and rewards to become more apparent. This method effectively balances the incorporation of recent data with the relevance of past information, thus refining our analysis of training episodes.

To assess the effectiveness of our proposed DRL approach, we compare the system performance with the following:
\begin{itemize}

    \item The $\epsilon$-greedy strategy, which is widely adopted in the literature to balance exploration and exploitation. 
    Using the $\epsilon$-greedy strategy, the CIoT agent selects an action aiming to maximize the estimated Q-value (exploitation) with a probability of $1-\epsilon$, while opting for a random action (exploration) with a probability of $\epsilon$. 
    \item The Fixed strategy involves the CIoT agent determining its action $a_t$ at each time step based on a rule-based approach derived from the constraints outlined in (\ref{eqn:optim}), without employing any learning mechanisms \cite{IoT_2024_Nada_Nadia}.
    \item The Random strategy entails the CIoT agent selecting an action $a_t$ at each time step randomly from the action space, without any cognitive or intelligent decision-making involved.
\end{itemize}
\subsection{Simulation Results}

\begin{figure}[t!]
    \centering
    \includegraphics[width = \columnwidth]{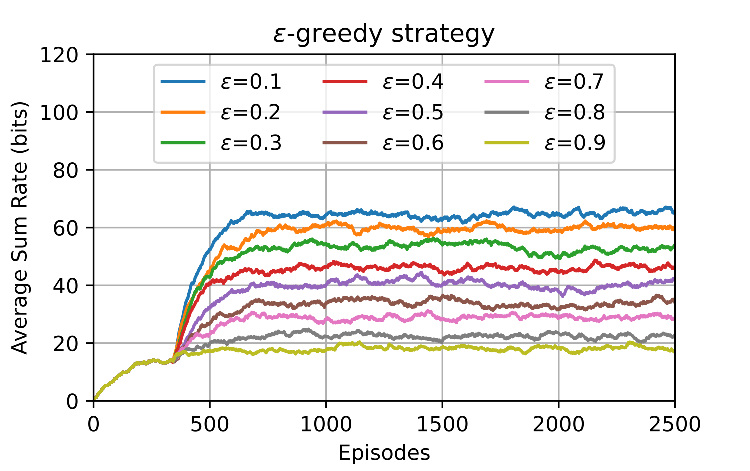}
    \caption{The CIoT Tx's ASR performance with $\epsilon$-greedy strategy across training episodes, comparison of different greediness value $\epsilon$.}
    \label{fig:thrpts_eps}
\end{figure}

To fairly compare our UCB-IA strategy with the $\epsilon$-greedy strategy, Fig.~\ref{fig:thrpts_eps} illustrates how different $\epsilon$ values affect the DRL algorithm's performance, allowing us to select the $\epsilon$ value that enhances the average sum rate (ASR). As the value of $\epsilon$ changes, the ASR of the CIoT agent changes considerably. When $\epsilon=$ 0.1, the highest ASR is observed. This suggests that lower $\epsilon$ values strike an optimal balance between exploration and exploitation, enabling the CIoT agent to make informed decisions based on its acquired knowledge while intermittently exploring new actions. Conversely, as $\epsilon$ increases, a decline in the CIoT Tx's ASR becomes evident. The lowest performance occurs at $\epsilon=$ 0.9, signifying that the CIoT agent heavily favors exploration and almost acts as the random strategy. This extensive exploration results in frequent penalties, as the CIoT agent randomly selects actions or disregards its learned knowledge, leading to suboptimal decisions and a decrease in overall performance. Therefore, we choose $\epsilon$= 0.1, which yields the best performance for the $\epsilon$-greedy strategy to ensure a fair comparison with our proposed approach.

\begin{figure}[t!]
    \centering
    \includegraphics[width = \columnwidth]{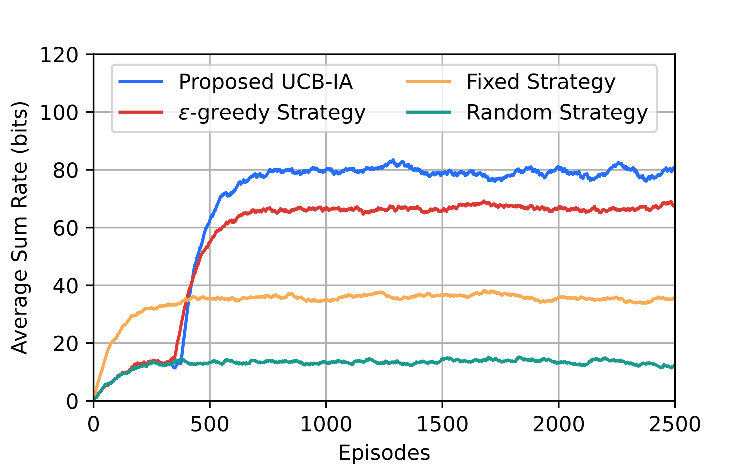}
    \caption{The CIoT Tx's ASR performance across training episodes, comparison of different strategies.}
    \label{fig:thrpts_benchmark}
\end{figure}

In Fig.~\ref{fig:thrpts_benchmark}, we present the ASR of the CIoT Tx across training episodes under several strategies. At the onset of training, both the proposed UCB-IA strategy and the $\epsilon$-greedy strategy attain an ASR similar to the random strategy, since the CIoT agent is in the process of accumulating experiences in the replay memory. However, at convergence, it is evident that our proposed UCB-IA strategy attains the highest ASR when compared with all other strategies. The CIoT agent's high performance with our DRL approach comes from the UCB-IA algorithm. This algorithm modifies the Q-value, helping the CIoT agent balance the trade-off between exploitation and exploration, considering both expected reward and jammer interference. Consequently, this allows the CIoT network to effectively share the spectrum with the primary network while maximizing its throughput under malicious jamming. 

The $\epsilon$-greedy strategy outperforms the random and fixed strategies, yet it falls significantly short of our proposed UCB-IA strategy, proving it is not the optimum approach to balance the exploitation-exploration trade-off in such a dynamic CIoT environment. The random strategy exhibits the lowest ASR, as its random action selection diminishes the likelihood of successful data transmission. The fixed strategy, adhering to predefined rules, ranks second-lowest in ASR, highlighting suboptimal action choices due to the CIoT agent's inability to explore all potential actions that can improve the ASR.

\begin{figure}[t!]
    \centering
    \includegraphics[width = \columnwidth]{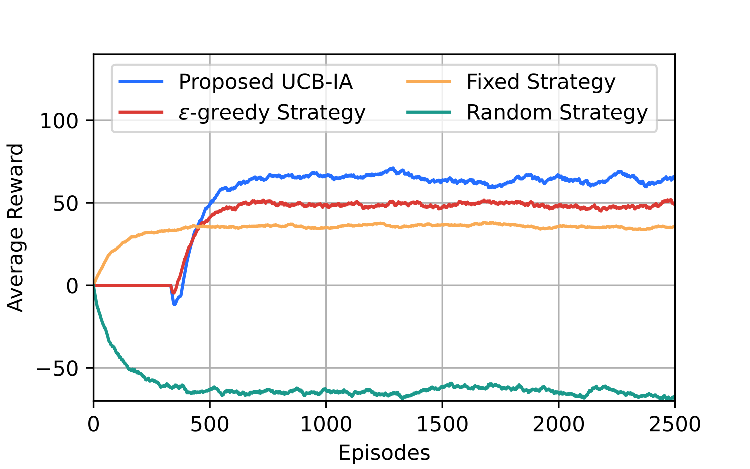}
    \caption{The CIoT Tx's average achievable reward across training episodes under different strategies.}
    \label{fig:rewards_benchmark}
\end{figure}

Fig.~\ref{fig:rewards_benchmark} illustrates the average rewards obtained by the CIoT agent across training episodes using various strategies. The figure demonstrates the convergence of all the learning-based strategies, affirming the effectiveness of the training approach. For both the proposed UCB-IA and $\epsilon$-greedy strategies, the reward remains zero at the beginning of training until the CIoT agent accumulates experiences in the replay buffer. Subsequently, both strategies exhibit a dip in rewards followed by an increase. This dip occurs because the DRL-based strategies favor long-term rewards over short-term rewards due to exploration. Similar to the trends observed in Fig.~\ref{fig:thrpts_benchmark}, our proposed DRL algorithm that employs the UCB-IA strategy attains the highest average reward, followed by the $\epsilon$-greedy strategy and the fixed strategy, while the random strategy records the lowest reward.

The fixed strategy consistently yields positive rewards throughout all training episodes due to its adherence to hard-coded rules, preventing actions that result in penalties. Conversely, the random strategy yields negative rewards in all episodes, attributed to its random action selection. Comparing Fig.~\ref{fig:thrpts_benchmark} and Fig.~\ref{fig:rewards_benchmark}, it can be noticed that for both UCB-IA and $\epsilon$-greedy, the ASR values are higher than the corresponding reward values. This is because the average reward metric accounts for penalties incurred during the exploration of new actions. However, despite receiving penalties for exploration, the reward of our proposed UCB-IA strategy remains higher than that of the $\epsilon$-greedy strategy.

\begin{figure}[t!]
    \centering
    \includegraphics[width = \columnwidth]{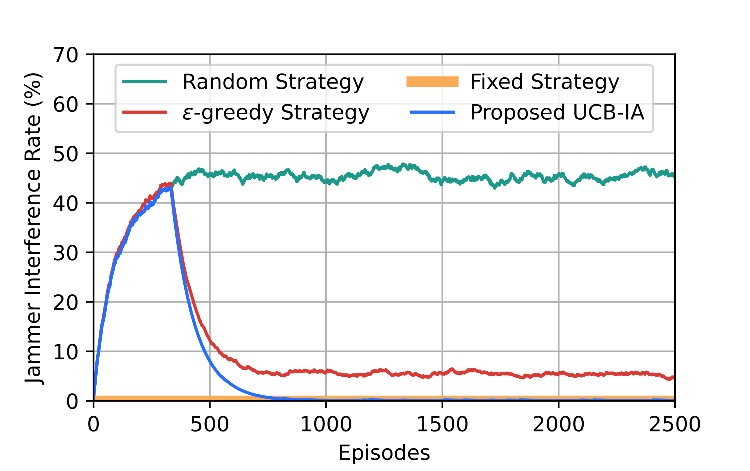}
    \caption{The jammer interference rate with the CIoT agent across training episodes under different strategies.}
    \label{fig:SU_interference}
\end{figure}

In Fig.~\ref{fig:SU_interference}, we depict the jammer interference rate across all training episodes for the four previously discussed strategies. Observations reveal that the random strategy exhibits the highest interference rate $\sim 50$\%. This suggests that, on average, the CIoT agent chooses to transmit data during approximately half of the time slots when jamming signals are present, leading to penalties and data loss. Although the fixed strategy ensures a zero interference rate for the CIoT agent throughout all training episodes, its lower performance in achieving the ASR, as depicted in Fig.~\ref{fig:thrpts_benchmark}, suggests that it is not the optimal method. The zero interference rate achieved by the fixed strategy is due to its rule-based approach, which adheres to the constraints outlined in equation (\ref{eqn:optim}), particularly constraint (\ref{constraint3}). This constraint dictates that the agent will switch to energy harvesting if the jammer is active during time slot $t$.

At the beginning of training, the jammer interference rate for the $\epsilon$-greedy and the UCB-IA strategy follows an increasing trend as the CIoT agent is in the process of accumulating experiences. However, as training progresses, the CIoT agent using the $\epsilon$-greedy strategy learns the actions that maximize the ASR, yet those actions ultimately result in an interference of 5\% with the jammer at convergence. In contrast, when the CIoT agent uses the proposed UCB-IA strategy, it effectively learns actions that not only maximize the long-term throughput but also result in a 0\% interference rate with the jammer. This implies that the UCB-IA strategy helps the CIoT device proficiently manage jammer interference on the same channel. It does so by harvesting from these interference signals, leading to higher battery levels and improved transmission success in future time slots.

\begin{figure}[t!]
    \centering
    \includegraphics[width = \columnwidth]{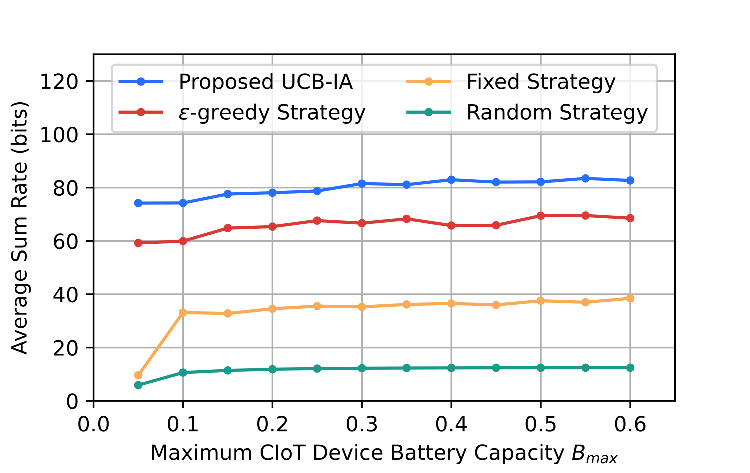}
    \caption{The effect of the maximum battery capacity $B_{max}$ of the CIoT agent on the ASR across different strategies.}
    \label{fig:thrpts_vs_B_Max}
\end{figure}

In Fig.~\ref{fig:thrpts_vs_B_Max}, we present the ASR for the four strategies across various values for maximum battery capacity $B_{max}$ of the CIoT device. As depicted in the figure, increasing $B_{\text{max}}$ allows the CIoT device to gather more energy through energy harvesting, resulting in increased ASR. Moreover, the increased capacity decreases the likelihood of battery overflow (when energy harvested exceeds $B_{\text{max}}$), thereby lowering the penalties faced by the CIoT agent. However, after reaching a certain threshold, any further increase in battery size offers minimal improvements, signifying a point of resource saturation.

At this stage, extra capacity fails to correspond to further performance enhancement. However, our proposed DRL algorithm that utilizes the UCB-IA strategy consistently outperforms all strategies across various values of $B_{max}$. This underscores the flexibility of our approach to optimize the ASR not only in scenarios with ample battery capacity but also in situations with limited capacity. It is important to highlight that, even at larger battery capacities, the performance of the CIoT agent employing the $\epsilon$-greedy strategy still remains inferior to our proposed approach.

\begin{figure}[t!]
    \centering
    \includegraphics[width = \columnwidth]{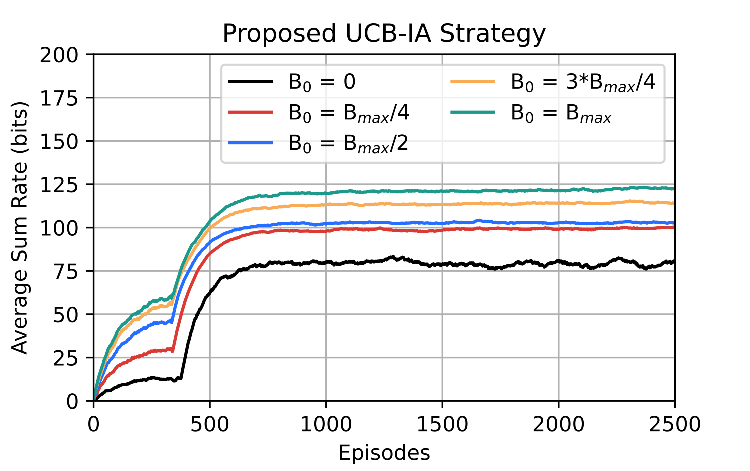}
    \caption{The effect of starting battery level $B_0$ on the ASR of the CIoT Tx using our proposed UCB-IA approach.}
    \label{fig:UCB_initial_battery_effect}
\end{figure}

In Fig.~\ref{fig:UCB_initial_battery_effect}, we present the ASR of the CIoT agent that uses our proposed DRL algorithm combined with the UCB-IA strategy across all training episodes, considering various initial battery levels $B_0$. Notably, the full battery configuration demonstrates the highest ASR compared to all other initial battery levels, while the empty battery configuration exhibits the lowest ASR. This is because, starting training with an empty battery limits the CIoT agent's available actions to achieve rewards, increasing the probability of choosing actions that result in penalties. Consequently, in such a scenario, the CIoT agent prioritizes energy harvesting in the early time slots to ensure the ability to transmit in subsequent time slots. Conversely, starting with a full battery allows the CIoT agent to prioritize data transmission, leading to higher rewards.

\begin{figure}[t!]
    \centering
    \includegraphics[width = \columnwidth]{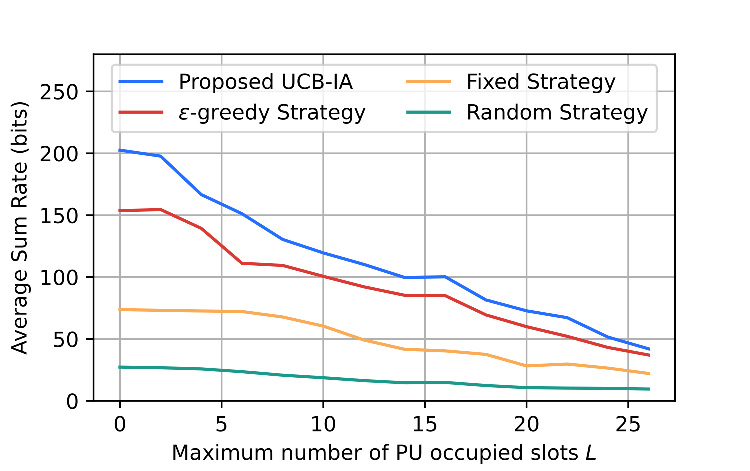}
    \caption{The effect of the number of PU transmission slots $L$ on the CIoT device’s ASR across different strategies.}
    \label{fig:thrpts_vs_numPU}
\end{figure}

In Fig.~\ref{fig:thrpts_vs_numPU}, we present the ASR of the CIoT agent under various spectrum-sharing scenarios by varying the number of slots occupied by the PU Tx $L$. Notably, as $L$ increases, there is a clear decrease in the ASR of all strategies. This decline can be attributed to the increased limitation on action selection by the CIoT agent as more occupied slots are introduced. This is because, during an occupied slot, the CIoT agent is restricted by the interference threshold $I_{th}$, necessitating the selection of lower transmit power $P_s^t$ to avoid penalties that result in a lower ASR. 

Moreover, it can be noticed that all strategies attain comparable performance at high values of $L$, which can be attributed to the fact that the PU Tx occupies nearly all the slots.  On the other hand, at low values of $L$, the CIoT agent has more flexibility in selecting its transmit power and, consequently, results in a higher ASR. Nevertheless, at various values of $L$, our proposed DRL algorithm utilizing the UCB-IA strategy consistently outperforms all other strategies. This shows that our proposed DRL approach can allow the CIoT network to achieve its goal of throughput maximization while efficiently coexisting with the PU Tx, even as channel occupancy changes.

\begin{figure}[t!]
     \centering
     \includegraphics[width = \columnwidth]{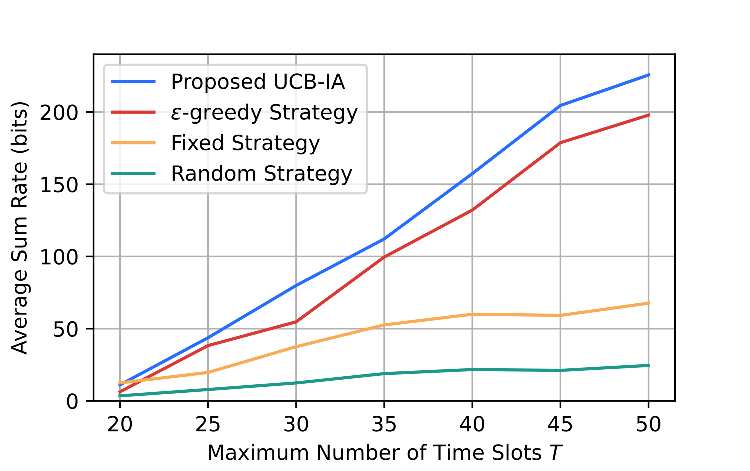}
     \caption{The effect of the number of time slots $T$ on the CIoT Tx’s ASR across different strategies.}
     \label{fig:Max_Time}
 \end{figure}

In Fig.~\ref{fig:Max_Time}, we present the ASR of the CIoT agent across various numbers of time slots $T$ under different strategies. It is evident that as the number of time slots increases, the ASR of all strategies increases. This is because, with a constant number of slots occupied by the PU Tx and targeted by the jammer, an increase in the number of time slots $T$ offers the CIoT agent more chances for transmission without facing penalties. As depicted in the figure, the proposed UCB-IA strategy and the $\epsilon$-greedy strategy demonstrate a more pronounced increase compared to the fixed and random strategies. 

This significant contrast can be attributed to their learning nature, which enables them not only to select the optimal action at a given time slot but also to determine the most advantageous transmit power, directly influencing the ASR. At 20 time slots, the ASR values for all four strategies are nearly identical. This is because at $T=20$ slots there are 18 slots occupied by the PU Tx and 10 slots subjected to jamming attacks. As a result, the available actions for the CIoT agent are restricted, resulting in all strategies converging on comparable choices. However, as $T$ increases, the ASR of the proposed UCB-IA strategy and the $\epsilon$-greedy strategy surpasses that of the fixed and random strategies by a considerable margin.

\section{Conclusions}

In this paper, we introduced a novel DRL algorithm designed to help CIoT networks effectively manage the exploitation-exploration trade-off in optimizing the action selection to achieve maximum ASR in malicious environments. This algorithm enables the CIoT agent to not only counteract jamming attacks but also utilize jamming interference to achieve its objectives and increase its lifetime. Our results show that the proposed DRL algorithm with the UCB-IA strategy meets its goals with considerable success, outperforming existing benchmarks significantly. This highlights the critical importance of tailoring DRL approaches to the specific system dynamics. We validated the algorithm's convergence under various network conditions, highlighting its potential to improve CIoT network performance even in challenging environments. These findings underscore the impact of learning algorithms in dynamic settings and suggest promising avenues for future research.

It is important to note that while our method excels in the specific context addressed in this paper, it may not always outperform existing benchmarks in simpler scenarios with fewer constraints. In such cases, our proposed strategy would achieve similar performance to existing benchmarks but would be less appealing due to its higher computational complexity and memory demands, making it more expensive to implement. Future research will focus on expanding the CIoT system model and investigating how network heterogeneity impacts performance. This expansion will require adjustments to the model and learning parameters, and the state and action spaces will become larger and more complex. To address these challenges and accelerate convergence, integrating transfer learning with DRL will be an interesting and promising direction for future work.

\bibliography{ref.bib}

@ARTICLE{9558789,  author={Ibrahim, Khalid and Ng, Soon Xin and Qureshi, Ijaz Mansoor and Malik, Aqdas Naveed and Muhaidat, Sami},  journal={IEEE Access},   title={{Anti-Jamming Game to Combat Intelligent Jamming for Cognitive Radio Networks}},   year={2021},  volume={9},  number={},  pages={137941-137956},  doi={10.1109/ACCESS.2021.3117563}}

@INPROCEEDINGS{eh_drl_sim_ppr,  author={Xie, Huan and Lin, Ruiquan and Wang, Jun and Zhang, Min and Cheng, Changchun},  booktitle={Proc. Int. Conf. Commun. Informat. Syst. (ICCIS)},   title={{Power Allocation of Energy Harvesting Cognitive Radio Based on Deep Reinforcement Learning}},   year={2021},  volume={},  number={},  pages={45-49},  doi={10.1109/ICCIS53528.2021.9645987}}

@ARTICLE{9606870,  author={Shi, Zhaoyuan and Xie, Xianzhong and Lu, Huabing and Yang, Helin and Cai, Jun and Ding, Zhiguo},  journal={IEEE Trans. Commun.},   title={{Deep Reinforcement Learning-Based Multidimensional Resource Management for Energy Harvesting Cognitive NOMA Communications}},   year={2022},  volume={70},  number={5},  pages={3110-3125},  doi={10.1109/TCOMM.2021.3126626}}

@ARTICLE{pirayesh2022jamming,
  author={Pirayesh, Hossein and Zeng, Huacheng},
  journal={IEEE Commun. Surveys Tuts.}, 
  title={{Jamming Attacks and Anti-Jamming Strategies in Wireless Networks: A Comprehensive Survey}}, 
  year={2022},
  volume={24},
  number={2},
  pages={767-809},
  doi={10.1109/COMST.2022.3159185}}

@book{bishop2006pattern,
  title={Pattern recognition and machine learning},
  author={Bishop, Christopher M and Nasrabadi, Nasser M},
  volume={4},
  number={4},
  year={2006},
  publisher={Springer}
}

@INPROCEEDINGS{nada_icc22,  author={A. Khalek, Nada and Hamouda, Walaa},  booktitle={Proc. IEEE Int. Conf. Commun. (ICC)},   title={{Intelligent Spectrum Sensing: An Unsupervised Learning Approach Based on Dimensionality Reduction}},   year={2022},  volume={},  number={},  pages={171-176},  doi={10.1109/ICC45855.2022.9839170}}

@INPROCEEDINGS{chen_power_control,
  author={Chen, Xiping and Xie, Xianzhong and Shi, Zhaoyuan and Fan, Zishen},
  booktitle={Proc. IEEE Int. Conf. Commun. China (ICCC)}, 
  title={{Dynamic Spectrum Access Scheme of Joint Power Control in Underlay Mode Based on Deep Reinforcement Learning}}, 
  year={2020},
  volume={},
  number={},
  pages={536-541},
  doi={10.1109/ICCC49849.2020.9238884}}

@ARTICLE{nada_survey_2023,
  author={Khalek, Nada Abdel and Tashman, Deemah H. and Hamouda, Walaa},
  journal={IEEE Commun. Surveys Tuts.}, 
  title={{Advances in Machine Learning-Driven Cognitive Radio for Wireless Networks: A Survey}}, 
  year={2024},
  volume={26},
  number={2},
  pages={1201-1237},
  doi={10.1109/COMST.2023.3345796}}

@ARTICLE{chu_eh_ciot_ma_2019,
  author={Chu, Man and Liao, Xuewen and Li, Hang and Cui, Shuguang},
  journal={IEEE Internet Things J.}, 
  title={{Power Control in Energy Harvesting Multiple Access System With Reinforcement Learning}}, 
  year={2019},
  volume={6},
  number={5},
  pages={9175-9186},
  doi={10.1109/JIOT.2019.2928837}}

@INPROCEEDINGS{He2015,
  author={He, Kaiming and Zhang, Xiangyu and Ren, Shaoqing and Sun, Jian},
  booktitle={Proc. IEEE Int. Conf. on Comput. Vis. (ICCV)}, 
  title={{Delving Deep into Rectifiers: Surpassing Human-Level Performance on ImageNet Classification}}, 
  year={2015},
  volume={},
  number={},
  pages={1026-1034},
  doi={10.1109/ICCV.2015.123}}

@inproceedings{Nguyen_Nguyen_Do_2021, title={{A Deep Double-$Q$ Learning-based Scheme for Anti-Jamming Communications}}, ISSN={2076-1465}, DOI={10.23919/Eusipco47968.2020.9287318}, abstractNote={Cognitive radio has become an emerging advanced wireless communication technology to achieve maximal spectrum efficiency. In cognitive radio networks, the threat of radio jamming attack arises as a big issue due to the vulnerability of radio transmission. Therefore, anti-jamming is an active research topic for a long time. Recently, with the success of deep learning, deep reinforcement learning algorithms have been applied to solve the dynamic spectrum access and anti-jamming problem. In this paper, we propose a Deep Double-Q learning-based method to learn an efficient communication policy including channel access and transmission power for tackling different jamming scenarios. The proposed scheme uses observed spectral information as input and Q-function is approximated by a neural network. Simulation results show that Double-Q learning algorithm with Convolutional Neural Network achieves effective communication strategies to avoid various jamming patterns compared with other traditional methods.}, booktitle={Proc. Eur. Signal Proc. Conf. (EUSIPCO)}, author={Nguyen, Phan Khanh Ha and Nguyen, Viet Hung and Do, Van Long}, year={2021}, month=jan, pages={1566-1570} }

@ARTICLE{8558506,
  author={Jia, Luliang and Xu, Yuhua and Sun, Youming and Feng, Shuo and Anpalagan, Alagan},
  journal={IEEE Wireless Commun.}, 
  title={{Stackelberg Game Approaches for Anti-Jamming Defence in Wireless Networks}}, 
  year={2018},
  volume={25},
  number={6},
  pages={120-128},
  doi={10.1109/MWC.2017.1700363}}

@inproceedings{Chen_Li_Xu_Xiao_2018, title={{DQN-Based Power Control for IoT Transmission against Jamming}}, ISSN={2577-2465}, DOI={10.1109/VTCSpring.2018.8417695}, abstractNote={Internet of Things (IoTs) have to address jammers, with goal to interrupt the communication of the energy- constrained IoT devices and sometimes even cause denial-of-service attacks. In this paper, we propose a deep reinforcement learning based power control scheme for IoT devices to improve the transmission efficiency and save energy. This scheme depends on the current IoT transmission status and the jamming strength and applies deep Q-network (DQN) to determine the transmit power without being aware of the IoT topology and the jamming model. This scheme is implemented on the universal software radio peripherals for the anti- jamming communication performance evaluation. Experimental results show that this scheme improves the signal-to-interference-plus-noise of the IoT signals compared with the benchmark Q-learning based power control scheme against jamming.}, booktitle={Proc. IEEE Veh. Technol. Conf. (VTC Spring)}, author={Chen, Ye and Li, Yanda and Xu, Dongjin and Xiao, Liang}, year={2018}, month=jun, pages={1-5} }

@article{Xu_Lou_Zhang_Sang_2020, title={{An Intelligent Anti-Jamming Scheme for Cognitive Radio Based on Deep Reinforcement Learning}}, volume={8}, ISSN={2169-3536}, DOI={10.1109/ACCESS.2020.3036027}, abstractNote={Cognitive radio network is an intelligent wireless communication system which can adjust its transmission parameters according to the environment thanks to its learning ability. It is a feasible and promising direction to solve the spectrum scarcity issue and has become a research focus in communication community. However, cognitive radio network is vulnerable to jamming attack, resulting in serious degradation of spectrum utilization. In this article, we view the anti-jamming task of cognitive radio as a Markov decision process and propose an intelligent anti-jamming scheme based on deep reinforcement learning. We aim to learn a policy for users to maximize their rate of successful transmission. Speciﬁcally, we design Double Deep Q Network (Double DQN) to model the confrontation between the cognitive radio network and the jammer. The Q network is implemented using Transformer encoder to effectively estimate action-values from raw spectrum data. The simulation results indicate that our approach can effectively defend against several kinds of jamming attacks.}, journal={IEEE Access}, author={Xu, Jianliang and Lou, Huaxun and Zhang, Weifeng and Sang, Gaoli}, year={2020}, pages={202563-202572}, language={en} }

@ARTICLE{8094301,
  author={Ahmed, Ismail K. and Fapojuwo, Abraham O.},
  journal={IEEE Trans. Cogn. Commun. Netw.}, 
  title={{Stackelberg Equilibria of an Anti-Jamming Game in Cooperative Cognitive Radio Networks}}, 
  year={2018},
  volume={4},
  number={1},
  pages={121-134},
  doi={10.1109/TCCN.2017.2769121}}

@INPROCEEDINGS{Ali_Naser_Muhaidat_2023,
  author={Ali, Abubakar S. and Naser, Shimaa and Muhaidat, Sami},
  booktitle={Proc. IEEE Int. Symp. Pers. Indoor Mob. Radio Commun. (PIMRC)}, 
  title={{Defeating Proactive Jammers Using Deep Reinforcement Learning for Resource-Constrained IoT Networks}}, 
  year={2023},
  volume={},
  number={},
  pages={1-6},
  doi={10.1109/PIMRC56721.2023.10293793}}

@article{Wang_Wang_Xu_Chen_Jia_Liu_Yang_2020, title={{Dynamic Spectrum Anti-Jamming Communications: Challenges and Opportunities}}, volume={58}, ISSN={1558-1896}, DOI={10.1109/MCOM.001.1900530}, abstractNote={Due to the openness of the transmission medium, it is necessary for radio systems to have anti-jamming abilities. Traditional anti-jamming methods such as sequence-based frequency hopping and direct sequence spread spectrum have shortcomings of low spectral efficiency and fixed communication patterns. With the development of software-defined radio, jamming devices are increasingly advanced and efficient. In this article, we propose a new paradigm for anti-jamming called DSAJ. With the help of cognitive radio and machine learning, the aim of DSAJ is to learn the dynamic and complex spectrum environment and obtain an optimal communication strategy. We first introduce the basic concept of anti-jamming communications and provide a brief summary of anti-jamming methods. Then, through a case study, mathematical modeling and applications of DSAJ are discussed for both single-user and multi-user systems. A real-life DSAJ testbed is described, and some potential research directions are discussed.}, number={2}, journal={IEEE Commun. Mag.}, author={Wang, Ximing and Wang, Jinlong and Xu, Yuhua and Chen, Jin and Jia, Luliang and Liu, Xin and Yang, Yijun}, year={2020}, month=feb, pages={79-85} }

@inproceedings{Shen_Wang_Jin_Zhang_2021, title={{Defend Against Jamming Attacks Using Deep Reinforcement Learning}}, ISSN={2643-4687}, DOI={10.1109/ISAPE54070.2021.9753520}, abstractNote={Cognitive radio network has the learning ability to adjust transmission behavior to adapt to dynamic electromagnetic environment, and it has become a research hotspot in the field of the new generation of intelligent wireless communication systems. However, recent research results show that cognitive radio networks are vulnerable to interference attacks, resulting in a serious decline in spectrum utilization. In this paper, an intelligent anti-jamming scheme based on deep reinforcement learning is proposed, and the anti-jamming decision of the communication party is modeled as a Markov decision chain. This method can help the communication transmitter to learn a combination of communication parameters that can maximize the probability of communication success. In the specific implementation, we use Double DQN as the communication decision-making brain, and specially designed a convolutional neural network suitable for the communication spectrum as input to accurately estimate the action value function. The simulation results show that our method has better performance than the traditional anti-jamming method based on reinforcement learning.}, booktitle={Proc. Int. Symp. Antennas Propag. EM Theory (ISAPE)}, author={Shen, Weiguo and Wang, Wei and Jin, Hu and Zhang, Weifeng}, year={2021}, month=dec, pages={1-3} }

@inproceedings{Liu_Li_Cui_Liu_Chen_Chen_Li_Xu_2023, title={{Flexible Channel Access Against Unknown Dynamic Jamming Attack: A Reinforcement Learning Approach}}, ISSN={2377-8644}, DOI={10.1109/ICCC57788.2023.10233388}, abstractNote={This paper investigates multi-domain anti-jamming communication in a malicious jamming and channel fading environment using deep reinforcement learning (DRL). Unlike existing research, which mainly focuses on single-domain anti-jamming methods such as frequency selection and power control, multi-domain anti-jamming is considered to cope with dynamic jamming and channel fading in the cognitive radio communication scene. We consider whether to transmit at the current time slot in the time domain and joint channel-bandwidth selection in the frequency domain. To represent the environment state, A two-dimensional time-frequency spatial-spectral matrix is designed. The channel switching and bandwidth adjustment overhead are considered in the instant reward to avoid frequent channel switching and bandwidth adjustment. We propose an intelligent anti-jamming algorithm based on DRL and long short term memory network (LSTM) layer is introduced into the DRL network. Simulation results demonstrate that our proposed algorithm is adaptive in dynamic jamming environments and outperforms other anti-jamming algorithms.}, booktitle={Proc. IEEE Int. Conf. Commun. China (ICCC)}, author={Liu, Songyi and Li, Wen and Cui, Li and Liu, Xin and Chen, Xueqiang and Chen, Runfeng and Li, Yangyang and Xu, Yuhua}, year={2023}, month=aug, pages={1-6} }

@article{Han_Xu_Jin_Li_Chen_Fang_Xu_2022, title={{Primary-User-Friendly Dynamic Spectrum Anti-Jamming Access: A GAN-Enhanced Deep Reinforcement Learning Approach}}, volume={11}, ISSN={2162-2345}, DOI={10.1109/LWC.2021.3125337}, abstractNote={This letter studies the problem of deep reinforcement learning (DRL)-based dynamic spectrum anti-jamming access in overlay cognitive radio networks. To prevent secondary user (SU) from interfering with primary user (PU) and being jammed by jammer, we propose a PU-friendly dynamic spectrum anti-jamming access scheme. First, a generative adversarial network (GAN)-based virtual environment is proposed to simulate spectrum environment. Then, a DRL-based channel decision network (CDN) is trained to learn the optimal spectrum access policy in the virtual environment. Finally, SU accesses spectrum environment under the guidance of the trained CDN. Simulation results show that the proposed scheme is able to elude both PU signals and jamming completely and converges much faster than the scheme that trains the CDN in spectrum environment from scratch.}, number={2}, journal={IEEE Wireless Commun. Lett.}, author={Han, Hao and Xu, Yifan and Jin, Zhu and Li, Wen and Chen, Xueqiang and Fang, Gui and Xu, Yuhua}, year={2022}, month=feb, pages={258-262} }

@inproceedings{Bi_Wu_Hua_2019, title={{Deep Reinforcement Learning Based Multi-User Anti-Jamming Strategy}}, ISSN={1938-1883}, DOI={10.1109/ICC.2019.8761848}, abstractNote={The threat of radio frequency jamming attack to cognitive radio network is an issue that has been discussed for a long time. Q-learning is a widely used anti-jamming algorithm due to its model-free characteristic. However, the traditional Q-learning based anti-jamming algorithms suffer from some limitations when dealing with high-dimensional or continuous inputs. The recently proposed double Deep Q-learning Network (DQN) overcomes this weakness by approximating the table based Q function with a deep neural network. In this paper, we apply the double DQN algorithm with frequency hopping strategy against RF jamming attack in a multi-user environment. We test the performances of three types of neural networks which are the fully connected network (FCN), the convolutional neural network (CNN) and the long short term memory (LSTM). The simulation shows the effectiveness of the double DQN algorithm. Meanwhile, the FCN agent gives the best result concerning stability.}, booktitle={Proc. IEEE Int. Conf. Commun. (ICC)}, author={Bi, Yue and Wu, Yue and Hua, Cunqing}, year={2019}, month=may, pages={1-6} }

@inproceedings{Aref_Jayaweera_2019, title={{Multi-task Deep Reinforcement Learning for Cognitive Spectrum-agile Communications}}, ISSN={2164-7011}, DOI={10.1109/ICIIS47346.2019.9063307}, abstractNote={This paper introduces a cognitive engine design to achieve spectrum-agile communications over a heterogeneous wideband spectrum. The proposed cognitive approach has the ability to learn and avoid interference signals and other harmful signals. The targeted spectrum in this work is much wider than the ones proposed in the literature, most likely covering several hundreds of MHz. The proposed approach is based on deep reinforcement learning (DRL), more specifically on a double deep Q-network (DDQN) made of a convolutional neural network (CNN). The wideband spectrum is divided into a number of sub-bands and each sub-band consists of a number of channels. The problem is modeled as a multi-task DRL, where each sub-band represents a single task. Transfer learning is used between tasks to speed up the learning process. It is shown, through simulations, that the proposed technique can efficiently learn an effective strategy to avoid harmful signals in a noncontiguous wideband spectrum. Furthermore, it outperforms other DRL-based approaches in the literature while operating in a much wider spectrum and maintaining low computational complexity.}, booktitle={Proc. Conf. Ind. Inf. Syst. (ICIIS)}, author={Aref, Mohamed A. and Jayaweera, Sudharman K.}, year={2019}, month=dec, pages={284-289} }

@inproceedings{Zhou_Li_Niu_Qin_Zhao_Wang_2020, title={{One Plus One is Greater Than Two: Defeating Intelligent Dynamic Jamming with Collaborative Multi-agent Reinforcement Learning}}, DOI={10.1109/ICCC51575.2020.9345127}, abstractNote={In this paper, we investigate the problem of anti-jamming communication in multi-user scenarios. The Markov game framework is introduced to model and analyze the anti-jamming problem, and a joint multi-agent anti-jamming algorithm (JMAA) is proposed to obtain the optimal anti-jamming strategy. In intelligent dynamic jamming environment, the JMAA adopts multi-agent reinforcement learning (MARL) to make on-line channel selection, which can effectively tackle the external malicious jamming and avoid the internal mutual interference among users. The simulation results show that the proposed JMAA is superior to the frequency-hopping based method, the sensing-based method and the independent Q-learning method.}, booktitle={Proc. IEEE Int. Conf. Comput. Commun. (ICCC)}, author={Zhou, Quan and Li, Yonggui and Niu, Yingtao and Qin, Zichao and Zhao, Long and Wang, Junwei}, year={2020}, month=dec, pages={1522-1526} }

@article{Li_Xu_Xu_Liu_Wang_Li_Anpalagan_2020, title={{Dynamic Spectrum Anti-Jamming in Broadband Communications: A Hierarchical Deep Reinforcement Learning Approach}}, volume={9}, ISSN={2162-2345}, DOI={10.1109/LWC.2020.2999333}, abstractNote={In this letter, the frequency selection problem in jamming environment with large number of optional frequencies is investigated. With numerous optional actions in the wider frequency band scenario, most of existing anti-jamming methods will become ineffective, since the convergence time and computational complexity will grow exponentially with the number of actions. To cope with the above challenge, a novel hierarchical deep reinforcement learning algorithm which does not need to know the jamming patterns and channel model is proposed. The proposed algorithm divides the frequency selection problem in the broadband into two steps via two subnetworks: Firstly, the frequency band is selected by the band selection network, and then the specific frequency is selected in this frequency band by the frequency selection network. Simulation results show that the proposed algorithm avoids multiple different jammings effectively and achieves satisfactory throughput with less calculation.}, number={10}, journal={IEEE Wireless Commun. Lett.}, author={Li, Yangyang and Xu, Yuhua and Xu, Yitao and Liu, Xin and Wang, Ximing and Li, Wen and Anpalagan, Alagan}, year={2020}, month=oct, pages={1616-1619} }

@article{Aref_Jayaweera_2021, title={{Spectrum-Agile Cognitive Radios Using Multi-Task Transfer Deep Reinforcement Learning}}, volume={20}, ISSN={1558-2248}, DOI={10.1109/TWC.2021.3076180}, abstractNote={This work proposes a cognitive engine design that enables a radio to find transmission opportunities in non-contiguous wideband spectrum to avoid interference. The radio’s objective is to apply both frequency hopping and transmit power adjustment to maintain a required level of quality-of-service (QoS). The spectrum is partitioned into sub-bands each made of a number of narrowband channels. A multi-task deep Q-network (DQN) is utilized to solve the underlying problem where communications over each sub-band represents a single task. The proposed technique exploits transfer learning between tasks to speed up learning operation for new tasks. The proposed multi-task transfer DQN is proved to be converged. It is shown through simulations that the radio is able to learn an efficient strategy to evade interference signals in a partially observable environment. The experimental results indicate that the proposed approach offers up to 24% improvement to the percentage of successful communications when compared to other RL-based approaches found in existing literature.}, number={10}, journal={IEEE Trans. Wireless Commun.}, author={Aref, Mohamed A. and Jayaweera, Sudharman K.}, year={2021}, month=oct, pages={6729-6742} }

@inproceedings{Aref_Jayaweera_2019_int_av, title={{Robust Deep Reinforcement Learning for Interference Avoidance in Wideband Spectrum}}, DOI={10.1109/CCAAW.2019.8904887}, abstractNote={This paper presents a design of a cognitive engine for interference and jamming resilience based on deep reinforcement learning (DRL). The proposed scheme is aimed at finding the spectrum opportunities in a heterogeneous wideband spectrum. In this paper we discuss a specific DRL mechanism based on double deep Q-learning (DDQN) with a convolutional neural network (CNN) to successfully learn such interference avoidance operation over a wideband partially observable environment. It is shown, through simulations, that the proposed technique has a low computational complexity and significantly outperforms other techniques in the literature, including other DRL-based approaches.}, booktitle={Proc. IEEE Cogn. Commun. Aerosp. Appl. Workshop (CCAAW)}, author={Aref, Mohamed A. and Jayaweera, Sudharman K.}, year={2019}, month=jun, pages={1-5} }

@inproceedings{Ali_Lunardi_2022, title={{Deep Reinforcement Learning Based Anti-Jamming Using Clear Channel Assessment Information in a Cognitive Radio Environment}}, ISSN={2771-7402}, DOI={10.1109/CommNet56067.2022.9993858}, abstractNote={Jamming as a type of denial of service attack has proved to be destructive to communication systems. This paper investigates and implements an anti-jamming scheme in a dynamic jamming environment. In our study, we utilize the clear channel assessment (CCA) information available in the MAC layer of a standard IEEE wireless device. Consequently, we eliminate the need for additional equipment to obtain the raw spectrum information. This contrast existing works which need a priori knowledge of the jamming patterns or employ raw spectrum information. The CCA information of all available spectrum channels is utilized as input states to train a double deep q-network (DDQN) agent online to mitigate the effects of jamming. Numerical results show that the proposed anti-jamming approach is effective in different jamming scenarios.}, booktitle={Proc. Int. Conf. Adv. Commun. Technol. Netw. (CommNet)}, author={Ali, Abubakar S. and Lunardi, Willian T. and Bariah, Lina and Baddeley, Michael and Lopez, Martin Andreoni and Giacalone, Jean-Pierre and Muhaidat, Sami}, year={2022}, month=dec, pages={1-6} }

@article{Liu_Xia_Hu_Zheng_Zhang_2022, title={{Optimal Time Allocation for Energy Harvesting Cognitive Radio Networks with Multichannel Spectrum Sensing}}, volume={2022}, ISSN={1530-8669}, DOI={10.1155/2022/3940132}, abstractNote={From the perspective of time domain and frequency domain, we investigate the energy harvesting cognitive radio networks (EH-CRNs) with multichannel, where the secondary transmitter (ST) opportunistically accesses the licensed subchannels to transmit packets by consuming the harvested energy. To explore the spectrum holes and improve the lifetime of the EH-CRNs, the ST scavenges energy from the radio-frequency (RF) signal in the wide band during the energy harvesting (EH) phase and exploits the harvested energy for sequential sensing and packet transmission during the rest of the time slot. Under the energy constraint, the secondary throughput is improved by optimizing the time allocation among the EH phase, sensing phase, and transmission phase. We formulate the secondary throughput with respect to the durations of the three phases, prove the existence of the optimal time allocation, and discuss the secondary throughput in three cases of the EH-CRNs. Finally, numerical results validate the theoretical results about the secondary throughput and explore the impacts of key system parameters.}, journal={Wireless Commun. Mobile Comput.}, publisher={Hindawi}, author={Liu, Xiaoying and Xia, Ming and Hu, Ping and Zheng, Kechen and Zhang, Shubin}, editor={Liu, Lei}, year={2022}, month=aug, pages={3940132} }

@book{Sutton_Barto_2020, place={Cambridge, Massachusetts; London, England}, title={Reinforcement learning: An introduction}, publisher={The MIT Press}, author={Sutton, Richard S. and Barto, Andrew}, year={2020}}

@article{Guo_Yang_Zou_Qian_Zhu_Hanzo_2019, title={{Joint Optimization of Power Splitting and Beamforming in Energy Harvesting Cooperative Networks}}, volume={67}, ISSN={0090-6778, 1558-0857}, DOI={10.1109/TCOMM.2019.2945324}, number={12}, journal={IEEE Trans. Commun.}, author={Guo, Haiyan and Yang, Zhen and Zou, Yulong and Qian, Mujun and Zhu, Jia and Hanzo, Lajos}, year={2019}, month=dec, pages={8247-8257}, language={en} }

@ARTICLE{Umeonwuka_Adejumobi_Shongwe_Thokozani_2024,
  author={Umeonwuka, Obumneme Obiajulu and Adejumobi, Babatunde Segun and Shongwe, Thokozani},
  journal={IEEE Access}, 
  title={{Deep Learning-Assisted Energy Prediction Modeling for Energy Harvesting in Wireless Cognitive Radio Devices}}, 
  year={2024},
  volume={12},
  number={},
  pages={8700-8720},
  doi={10.1109/ACCESS.2023.3349352}}

@ARTICLE{Lin_Qiu_Wang_Zhang_2024,
  author={Lin, Ruiquan and Qiu, Hangding and Wang, Jun and Zhang, Zaichen and Wu, Liang and Shu, Feng},
  journal={IEEE Internet Things J.}, 
  title={{Physical-Layer Security Enhancement in Energy-Harvesting-Based Cognitive Internet of Things: A GAN-Powered Deep Reinforcement Learning Approach}}, 
  year={2024},
  volume={11},
  number={3},
  pages={4899-4913},
  doi={10.1109/JIOT.2023.3300770}}

@ARTICLE{9380308,
  author={Gouissem, Ala and Abualsaud, Khalid and Yaacoub, Elias and Khattab, Tamer and Guizani, Mohsen},
  journal={IEEE Internet Things J.}, 
  title={{Game Theory for Anti-Jamming Strategy in Multichannel Slow Fading IoT Networks}}, 
  year={2021},
  volume={8},
  number={23},
  pages={16880-16893},
  doi={10.1109/JIOT.2021.3066384}}

@ARTICLE{9172090,
  author={Xu, Yifan and Xu, Yuhua and Dong, Xu and Ren, Guochun and Chen, Jin and Wang, Ximing and Jia, Luliang and Ruan, Lang},
  journal={IEEE Trans. Veh. Technol.}, 
  title={{Convert Harm Into Benefit: A Coordination-Learning Based Dynamic Spectrum Anti-Jamming Approach}}, 
  year={2020},
  volume={69},
  number={11},
  pages={13018-13032},
  doi={10.1109/TVT.2020.3018121}}

@ARTICLE{9200488,
  author={Li, Yongcheng and Bai, Shaozhuang and Gao, Zhenzhen},
  journal={IEEE Access}, 
  title={{A Multi-Domain Anti-Jamming Strategy Using Stackelberg Game in Wireless Relay Networks}}, 
  year={2020},
  volume={8},
  number={},
  pages={173609-173617},
  doi={10.1109/ACCESS.2020.3025160}}

@ARTICLE{9099069,
  author={Zhou, Quan and Li, Yonggui and Niu, Yingtao},
  journal={IEEE Access}, 
  title={{A Countermeasure Against Random Pulse Jamming in Time Domain Based on Reinforcement Learning}}, 
  year={2020},
  volume={8},
  number={},
  pages={97164-97174},
  doi={10.1109/ACCESS.2020.2996804}}

@Article{s23020807,
AUTHOR = {Lin, Ruiquan and Qiu, Hangding and Jiang, Weibin and Jiang, Zhenglong and Li, Zhili and Wang, Jun},
TITLE = {{Deep Reinforcement Learning for Physical Layer Security Enhancement in Energy Harvesting Based Cognitive Radio Networks}},
JOURNAL = {Sensors},
VOLUME = {23},
YEAR = {2023},
NUMBER = {2},
ARTICLE-NUMBER = {807},
PubMedID = {36679601},
ISSN = {1424-8220},
DOI = {10.3390/s23020807}
}

@INPROCEEDINGS{9778818,
  author={Geng, Shuqin and Li, Pengkun and Yin, Xuzhou and Lu, Hang and Zhu, Ronghao and Cao, Wenhua and Nie, Jingyao},
  booktitle={Proc. Int. Conf. Intell. Comput. Signal Proc. (ICSP)}, 
  title={{The Study on Anti-Jamming Power Control Strategy based on $Q$-learning}}, 
  year={2022},
  volume={},
  number={},
  pages={182-185},
  doi={10.1109/ICSP54964.2022.9778818}}

@INPROCEEDINGS{10437464,
  author={{N. A. Khalek, and W. Hamouda}},
  booktitle={Proc. IEEE Glob. Commun. Conf. (GLOBECOM)}, 
  title={{DEAP Learning: A Data-Driven Approach to Unsupervised Cooperative Spectrum Sensing}}, 
  year={2023},
  volume={},
  number={},
  pages={6389-6394},
  doi={10.1109/GLOBECOM54140.2023.10437464}}

@ARTICLE{IoT_2024_Nada_Nadia,
  author={Khalek, Nada Abdel and Abdolkhani, Nadia and Hamouda, Walaa},
  journal={IEEE Internet Things J.}, 
  title={{Deep Reinforcement Learning for Joint Power Control and Access Coordination in Energy Harvesting CIoT}}, 
  year={2024},
  volume={},
  number={},
  pages={},
  doi={10.1109/JIOT.2024.3416371}}

@INPROCEEDINGS{10279156,
  author={Khalek, Nada Abdel and Hamouda, Walaa},
  booktitle={Proc. IEEE Int. Conf. Commun. (ICC)}, 
  title={{DeepSense: An Unsupervised Deep Clustering Approach for Cooperative Spectrum Sensing}}, 
  year={2023},
  volume={},
  number={},
  pages={1868-1873},
  doi={10.1109/ICC45041.2023.10279156}}

@ARTICLE{Tanab_survey_2017,
  author={El Tanab, Manal and Hamouda, Walaa},
  journal={IEEE Commun. Surveys Tuts.}, 
  title={{Resource Allocation for Underlay Cognitive Radio Networks: A Survey}}, 
  year={2017},
  volume={19},
  number={2},
  pages={1249-1276},
  doi={10.1109/COMST.2016.2631079}}
\bibliographystyle{IEEEtran}

\begin{IEEEbiography}[{\includegraphics[width=1in,height=1.25in,clip,keepaspectratio]{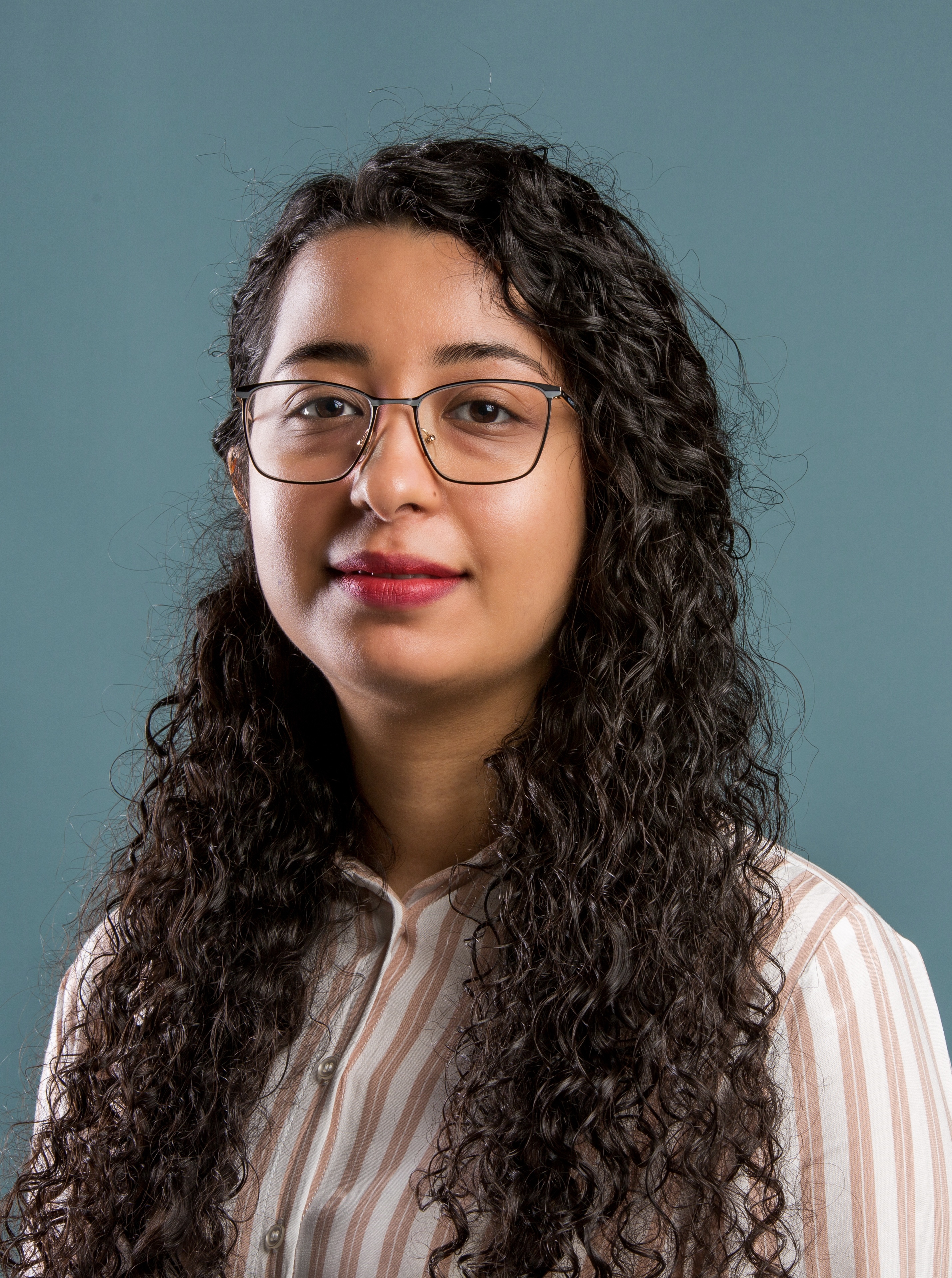}}]{Nadia Abdolkhani} 
(Student Member, IEEE) received the B.Sc. and M.Sc. degrees in electrical engineering from Shiraz University of Technology, Shiraz, Iran, in 2017 and 2021, respectively. Currently, she is pursuing a Ph.D. from the Department of Electrical and Computer Engineering at  Concordia University, Montreal, Canada. Her research interests include wireless and cellular communications, caching at wireless networks, device-to-device (D2D) communications, reinforcement learning-driven networks, and cognitive and cooperative communications.
\end{IEEEbiography}

\begin{IEEEbiography}[{\includegraphics[width=1in,height=1.25in,clip,keepaspectratio]{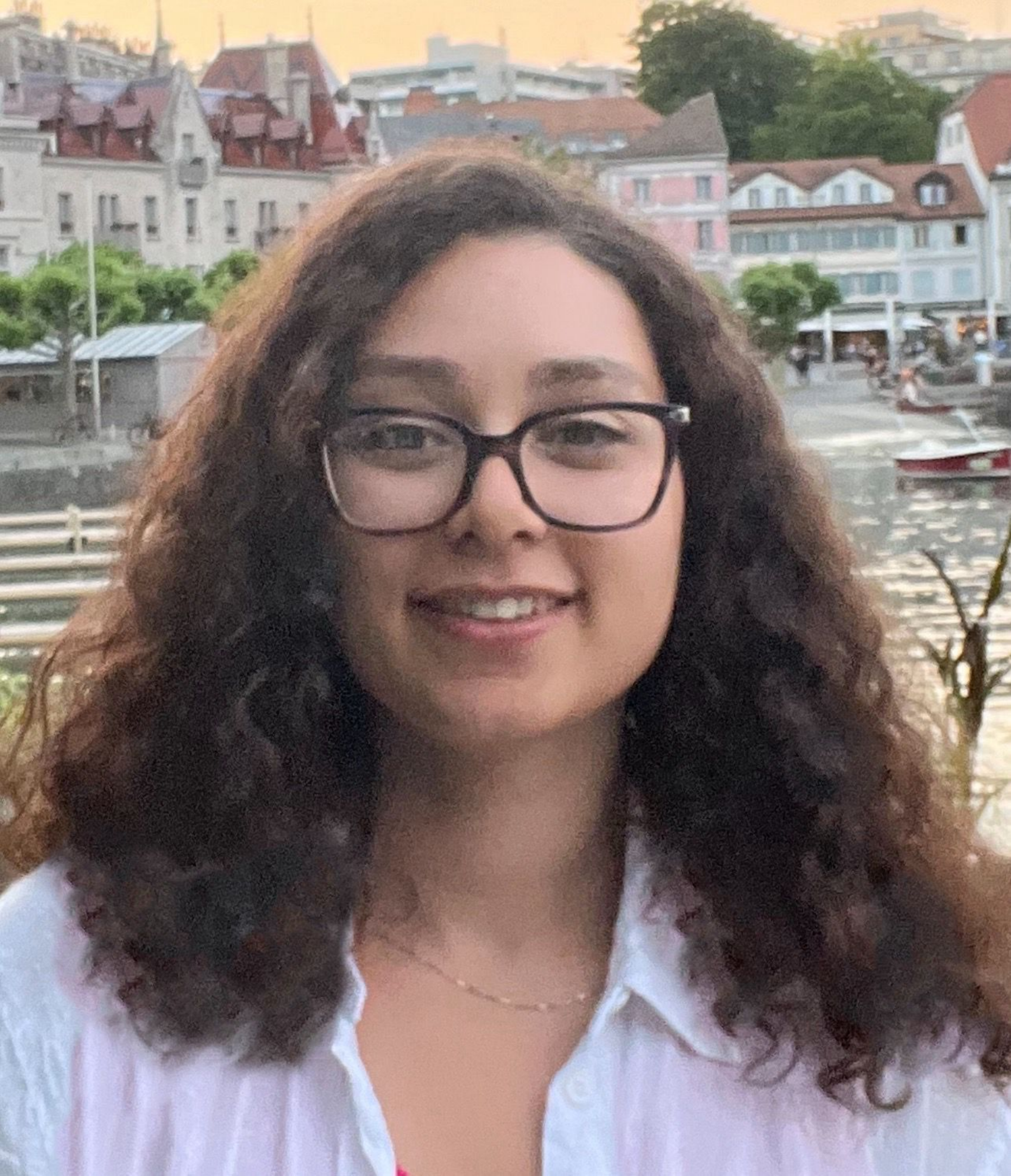}}]{Nada Abdel Khalek}
(Student Member, IEEE) received the M.Sc. degree in Electrical and Computer Engineering from Concordia University, Montreal, Quebec, Canada in 2020 and the B.Sc. degree (Magna Cum Laude) in Electronics and Communications Engineering from the American University in Cairo, Egypt, in 2018. She is currently pursuing her Ph.D. degree from the Department of Electrical and Computer Engineering at Concordia University. Her research focuses on intelligent wireless communications, including deep learning and reinforcement learning-driven network operations, cognitive and cooperative communications, and spectrum sharing. She is a recipient of both Concordia University's International Tuition Award of Excellence and Merit Scholarship. Additionally, she received the Fonds de Recherche du Quebec - Nature and Technologies (FRQNT) Doctoral Fellowship for 2022 and the Best Paper Award from IEEE Globecom 2023.
\end{IEEEbiography}

\begin{IEEEbiography}[{\includegraphics[width=1in,height=1.25in,clip,keepaspectratio]{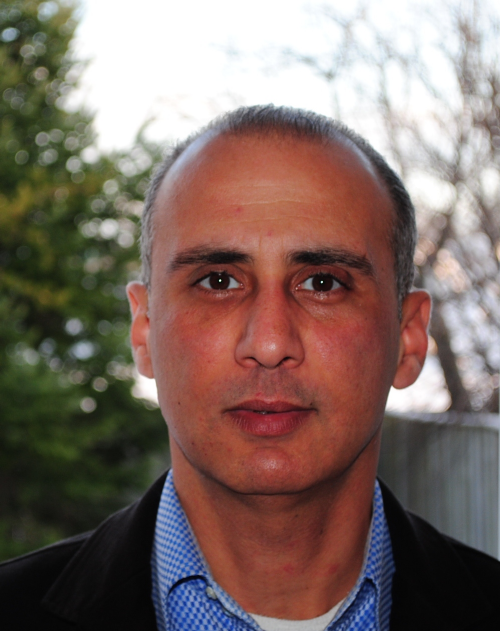}}]{Walaa Hamouda} (Senior Member, IEEE) received the M.A.Sc. and Ph.D. degrees in electrical and computer engineering from Queen’s University, Kingston, ON, Canada, in 1998 and 2002, respectively. Since July 2002, he has been with the Department of Electrical and Computer Engineering, Concordia University, Montreal, QC, Canada, where he is currently a Professor. Since June 2006, he is Concordia University Research Chair Tier I in Wireless Communications and Networking. His current research interests include machine-to-machine communications, IoT, 5G and beyond technologies, single/multiuser multiple-input multiple-output communications, space-time processing, cognitive radios, wireless networks. Dr. Hamouda served(ing) as Co-chair of the IoT and Sensor Networks Symposium of the GC’22, TPC Co-chair of the ITC/ADC 2022 conference, track Co-Chair: Antenna Systems, Propagation, and RF Design, IEEE Vehicular Technology Conference (VTC Fall'20), Tutorial Chair of IEEE Canadian Conference in Electrical and Computer Engineering (CCECE 2020), General Co-Chair, IEEE SmartNets 2019 Conference, Co-Chair of the MAC and Cross Layer Design Track of the IEEE (WCNC) 2019,  Co-chair of the Wireless Communications Symposium of the IEEE ICC’18, Co-chair of the Ad-hoc, Sensor, and Mesh Networking Symposium of the IEEE GC’17, Technical Co-chair of the Fifth International Conference on Selected Topics in Mobile \& Wireless Networking (MoWNet’2016), Track Co-Chair: Multiple Antenna and Cooperative Communications, IEEE Vehicular Technology Conference (VTC Fall'16), Co-Chair: ACM Performance Evaluation of Wireless Ad Hoc, Sensor, and Ubiquitous Networks (ACMPE-WASUN'14) 2014, Technical Co-chair of the Wireless Networks Symposium, 2012 Global Communications Conference, the Ad hoc, Sensor, and Mesh Networking Symposium of the 2010 ICC, and the 25th Queen’s Biennial Symposium on Communications. He also served as the Track Co-chair of the Radio Access Techniques of the 2006 IEEE VTC Fall 2006 and the Transmission Techniques of the IEEE VTC-Fall 2012. From September 2005 to November 2008, he was the Chair of the IEEE Montreal Chapter in Communications and Information Theory. He is an IEEE ComSoc Distingushed Lecturer. He received numerous awards, including the Best Paper Awards of the IEEE GC 2023, IEEE GC 2020, IEEE ICC 2021, ICCSPA 2019, IEEE WCNC 2016, IEEE ICC 2009, and the IEEE Canada Certificate of Appreciation in 2007 and 2008. He served as an Associate Editor for the IEEE COMMUNICATIONS LETTERS, IEEE TRANSACTIONS ON SIGNAL PROCESSING, IEEE COMMUNICATIONS SURVEYS AND TUTORIALS, IET WIRELESS SENSOR SYSTEMS, IEEE WIRELESS COMMUNICATIONS LETTERS, TRANSACTIONS ON VEHICULAR TECHNOLOGY, and currently serves as an Editor for the IEEE TRANSACTIONS ON COMMUNICATIONS, IEEE TRANSACTIONS ON WIRELESS COMMUNICATIONS, IEEE IoT journal.
\end{IEEEbiography}

\end{document}